\documentclass[iop,apj]{emulateapj}
\usepackage{amsmath,amssymb,amstext}

\usepackage[breaklinks,colorlinks,citecolor=blue,linkcolor=magenta]{hyperref}

\usepackage{url}

\usepackage{aas_macros}
\usepackage{natbib}
\def\farcsec{\hbox{$\ \!\!^{\prime\prime}$}}
\newcommand{\sm}{M_\odot}
\newcommand{\sr}{R_\odot}
\newcommand{\RNum}[1]{\uppercase\expandafter{\romannumeral #1\relax}}

\shorttitle{\lowercase{i}PTF13\lowercase{asv}}
\shortauthors{Cao et al.}

\begin{document}

\title{Absence Of Fast-Moving Iron In An Intermediate Type I\MakeLowercase{a} Supernova Between Normal And Super-Chandrasekhar}

\author{Yi~Cao\altaffilmark{1}, J.~Johansson\altaffilmark{2}, Peter~E.~Nugent\altaffilmark{3,4}, 
	A.~Goobar\altaffilmark{5}, Jakob~Nordin\altaffilmark{6}, S.~R.~Kulkarni\altaffilmark{1}, 
	S.~Bradley~Cenko\altaffilmark{7,8}, Ori~D.~Fox\altaffilmark{4,9}, Mansi~M.~Kasliwal\altaffilmark{1,10}, 
	C.~Fremling\altaffilmark{11}, R.~Amanullah\altaffilmark{5}, E.~Y.~Hsiao\altaffilmark{12,13}, D.~A.~Perley\altaffilmark{1}, 
	Brian~D.~Bue\altaffilmark{14}, Frank J. Masci\altaffilmark{15},
	William H. Lee\altaffilmark{16}, Nicolas~Chotard\altaffilmark{17}}
\altaffiltext{1}{Astronomy Department, California Institute of Technology, Pasadena, CA 91125, USA}
\altaffiltext{2}{Benoziyo Center for Astrophysics, Weizmann Institute of Science, 76100 Rehovot, Israel}
\altaffiltext{3}{Computational Cosmology Center, Computational Research Division, Lawrence Berkeley National Laboratory,
1 Cyclotron Road, MS 50B-4206, Berkeley, CA 94720, USA}
\altaffiltext{4}{Department of Astronomy, University of California Berkeley, Berkeley, CA 94720-3411, USA}
\altaffiltext{5}{Oskar Klein Centre, Physics Department, Stockholm University, SE-106 91 Stockholm, Sweden}
\altaffiltext{6}{Institut f\"{u}r Physik, Humboldt-Universit\"{a}t zu Berlin, Newtonstr. 15, 12489 Berlin, Germany}
\altaffiltext{7}{Astrophysics Science Division, NASA Goddard Space Flight Center, Mail Code 661, Greenbelt, Maryland 20771, USA}
\altaffiltext{8}{Joint Space-Science Institute, University of Maryland, College Park, MD 20742, USA}
\altaffiltext{9}{Space Telescope Science Institute, 3700 San Martin Drive, Baltimore, MD 21218, USA}
\altaffiltext{10}{Observatories of the Carnegie Institution for Science, 813 Santa Barbara Street, Pasadena, California 91101, USA}
\altaffiltext{11}{Department of Astronomy, The Oskar Klein Center, Stockholm University, AlbaNova, 10691 Stockholm, Sweden}
\altaffiltext{12}{Department of Physics, Florida State University, Tallahassee, FL 32306, USA}
\altaffiltext{13}{Department of Physics and Astronomy, Aarhus University, Ny Munkegade 120, 8000 Aarhus C, Denmark}
\altaffiltext{14}{Jet Propulsion Laboratory, California Institute of Technology, Pasadena, CA 91125, USA}
\altaffiltext{15}{Infrared Processing and Analysis Center, California Institute of Technology, MS 100-22, Pasadena, CA 91125, USA}
\altaffiltext{16}{Instituto de Astronom\'{\i}a, Universidad Nacional Aut\'{o}noma de M\'{e}xico, Apdo. Postal 70-264 Cd. Universitaria, 
M\'{e}xico DF 04510, M\'{e}xico}
\altaffiltext{17}{Universit\'e de Lyon, F-69622, France ; Universit\'e de Lyon 1, Villeurbanne ; CNRS/IN2P3, Institut de Physique Nucl\'eaire de Lyon}

\begin{abstract}
In this paper, we report observations of a peculiar Type \RNum{1}a supernova iPTF13asv (a.k.a., SN2013cv) from the onset of the explosion
to months after its peak. The early-phase spectra of iPTF13asv show absence of iron absorption, indicating that synthesized iron elements 
are confined to low-velocity regions of the ejecta, which, in turn, implies a stratified ejecta structure along the line of sight. Our analysis of iPTF13asv's 
light curves and spectra shows that it is an intermediate case between normal and super-Chandrasekhar events. On the one 
hand, its light curve shape (B-band $\Delta m_{15}=1.03\pm0.01$) and overall spectral features resemble those of normal Type \RNum{1}a 
supernovae. On the other hand, similar to super-Chandrasekhar events, it shows large peak optical and UV luminosity ($M_B=-19.84\,\rm{mag}$, $M_{uvm2}=-15.5\,\rm{mag}$)
a relatively low but almost constant \ion{Si}{2} velocities of about $10,000\,\rm{km}\,\rm{s}^{-1}$, 
and persistent carbon absorption in the spectra. 
We estimate a $^{56}$Ni mass of $0.81^{+0.10}_{-0.18}\sm$ and a total ejecta mass of $1.59^{+0.45}_{-0.12}\sm$. The large ejecta mass 
of iPTF13asv and its stratified ejecta structure together seemingly favor a double-degenerate origin. 
\end{abstract}

\keywords{supernovae: general -- supernovae: individual (iPTF13asv, SN2013cv) -- ultraviolet: general}
\maketitle

\section{Introduction}
\label{sec:introduction}
Type \RNum{1}a supernovae (SNe) are thermonuclear explosions of carbon-oxygen white dwarfs (WDs). Since the majority of them (the normal Type SNe Ia)
follow a well-established empirical relation between variation of their peak magnitudes and light curve shapes \citep{Phillips1993}, they are standardized
to measure cosmological distances (see \citealt{GL2011} for a review). However, the underlying progenitor systems and explosion mechanisms of
SNe Ia remain poorly understood. 

Recent observations have provided mounting evidence that SNe Ia have multiple progenitor channels (see \citealt{mmn14}
for a review). In the single-degenerate (SD) channel, a WD accretes material from a companion star and explodes when its mass approaches
the Chandrasekhar limit \citep{wi73}. This channel is supported by possible detections of companion stars in pre- or post-SN images
\citep{mjf+14,fmj+14}, likely signatures of SN-companion collisions \citep{ckh+15,mbv+16}, and observations of variable 
\ion{Na}{1}\,D absorption \citep{pcc+07,sgs+14}. In the double-degenerate (DD) channel, in contrast, two WDs collide or merge in 
a binary or even triple system to produce a SN \RNum{1}a \citep[e.g.,][]{ni85,kkd+13}. This channel is consistent with observations of 
two nearby Type \RNum{1}a SN2011fe and SN2014J \citep[e.g.,][]{lbp+11,BDd2012,ssp+13,mpk+14,kff+14,gks+15,lnt+15}. Despite
these interesting constraints from individual events, the progenitors of most SNe Ia are still unknown.

In the SD channel, rigid rotation may provide additional support for a WD of a mass slightly larger than the Chandrasekhar limit
and differential rotation may support for an even more massive WD. However, the theoretical viability of massive, rotation-supported WDs 
is much less clear in reality \citep{yl04,sn04,p08,j11,11dvc,hkn+12}.
In the DD channel, in contrast, the exploding WD binary may allow SN ejecta mass much higher than the Chandrasekhar limit.
In fact, more than a handful of SNe were found to have total ejecta masses significantly exceeding the Chandrasekhar limit
\citep{hsn+06,hgp+07,yqw+10,ykk+09,saa+10,sgl+11,sct+14}. However, these super-Chandrasekhar SNe show distinctive characteristics compared to normal
events: they are overluminous in both the optical and UV, implying a large amount of synthesized $^{56}$Ni. They show low expansion velocities and long
rise times, leading to massive ejecta. They also show persistent absorption from unburned carbon. 

In this paper, we present observations of a peculiar SN \RNum{1}a, iPTF13asv, which shares observational characteristics with both super-Chandrasekhar and normal
SN \RNum{1}a. It was discovered with $r=20.54\pm0.16$\,mag at $\alpha=16^h22^m43^s.19$, 
$\delta=+18^\circ57^\prime35\farcsec.0$ (J2000) in the vicinity of galaxy SDSS\,J$162243.02{+}185733.8$ on UTC 2013 May $1.44$ (hereafter May 
$1.44$) by the intermediate Palomar Transient Factory (iPTF; \citealt{LKD2009,RKL2009}). Nothing was
seen at the same location down to 5-$\sigma$ detection thresholds of $r\simeq 21.0$\,mag on images taken on April 30.5 and earlier. iPTF13asv was
independently discovered and classified as a peculiar type \RNum{1}a by \citet{CBET3543}, and was designated as SN2013cv. 

This paper is organized as follows: the observational data are presented in \S\ref{sec:obs}. The photometric and spectroscopic 
properties are analyzed in \S\ref{sec:analysis}, and we construct its bolometric light curve and estimate the total ejecta mass in \S\ref{sec:mass}. 
A discussion of the nature of iPTF13asv is given in \S\ref{sec:discussion} and our conclusions are summarized in \S\ref{sec:conclusion}. 

In order to have a comparison to other SNe, we adopt a fiducial value of the Hubble constant 
$H_0=72\,{\rm km}\,{\rm s}^{-1}\,{\rm Mpc}^{-1}$. The apparent host galaxy of iPTF13asv does not have a redshift-independent 
distance measurement in the NASA/IPAC Extragalactic Database. Thus the redshift $0.036$ leads to a distance modulus of 35.94\,mag. 
The peculiar motion of the host galaxy at ${\sim}100\,\rm{km}\,\rm{s}^{-1}$ introduces an uncertainty of $\lessapprox0.05$\,mag to the distance modulus. 
The Galactic line-of-sight extinction is $E(B-V)=0.045$ \citep{SF2011}. We correct for the Galactic extinction by using the parameterized model in
\citet{ftz99} with $R_V=3.1$.

\section{Observations}
\label{sec:obs}
The nightly cadence survey of iPTF (weather permitting) with the 48-inch telescope at the Palomar Observatories (P48)
provides a well-sampled R-band light curve of iPTF13asv covering the pre-SN history and its rise phase. After 
discovery, we also utilized the Palomar 60-inch telescope (P60; \citealt{CFM2006}), the Andalusia Faint Object Spectrograph and 
Camera (ALFOSC) on the Nordic Optical Telescope (NOT), and the RATIR camera mounted on the OAN/SPM $1.5$-meter Harold L. Johnson 
telescope for multi-band photometric follow-up observations. We also triggered target-of-opportunity observations of the \textit{Swift}
spacecraft for X-ray and UV follow-up. The ground-based photometric measurements are presented in Table~\ref{tab:lightcurve}
and the space-based measurements in Table~\ref{tab:swift}. 

\begin{deluxetable}{ccccc}
    \tablecolumns{5}
    \tablewidth{0pt}
    \tablecaption{Photometry of iPTF13asv \label{tab:lightcurve}}
    \tablehead{
        \colhead{Tel./Inst.\tablenotemark{1}}  &  \colhead{Filter}  &  \colhead{MJD$-56000$}  &  \colhead{mag.\tablenotemark{2}}  &  \colhead{mag. err.} \\
                         &  \colhead{(day)}      &                    &                  &                  }
    \startdata
        P48   &    PTF-R      & 412.456      &    $>$20.35    &   \nodata\\
        P48   &    PTF-R      & 413.442      &     20.44    &   0.17\\
        P48   &    PTF-R      & 413.471      &     20.34    &   0.16\\
        P48   &    PTF-R      & 414.470      &     19.58    &   0.09\\
        P48   &    PTF-R      & 415.469      &     19.06    &   0.06\\
        \multicolumn{5}{c}{\nodata} 
    \enddata
    \tablenotetext{1}{This column lists the telescopes and instruments used for photometric observations of iPTF13asv. We built reference images
    by stacking pre-SN or post-SN frames and used an image subtraction technique to remove light contamination from the host galaxy. Point-spread function 
    (PSF) photometry is performed on subtracted images. The photometry is calibrated either to SDSS or by observing Landolt photometric standard stars. 
    }
    \tablenotetext{2}{Conventionally, the magnitudes in $UBVJH$ bands are in the Vega system. Those in other bands are in 
    the AB system. No extinction is corrected in this column. }
\end{deluxetable}

\begin{figure}[htb]
\centering
\includegraphics[width=0.45\textwidth]{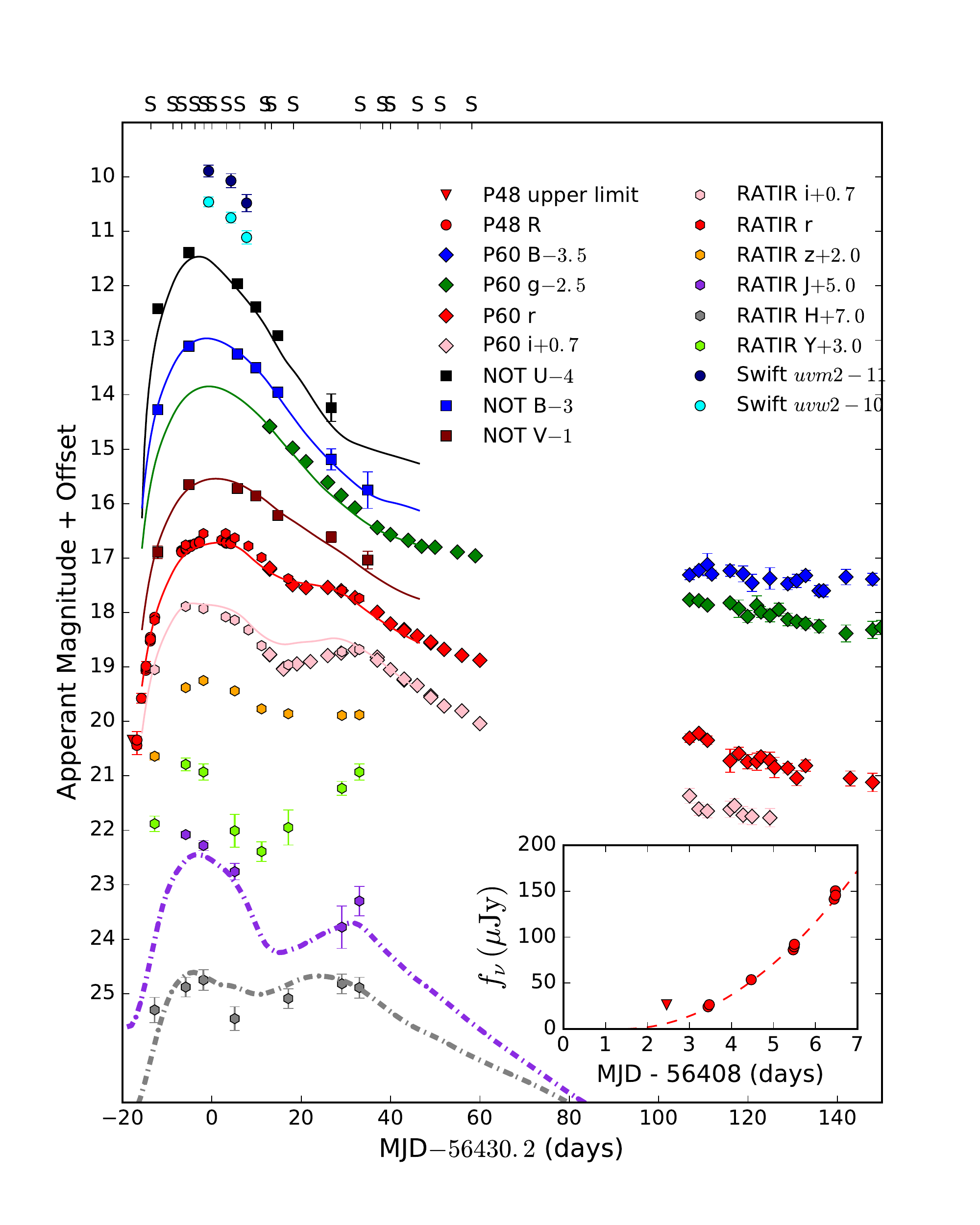}
\caption{Multi-color light curve of iPTF13asv Colors and shapes represent different filters and instruments, respectively. 
A deviation of $\simeq0.1$\,mag between the P48 R-band (red circles) and P60 r-band (red diamonds) is due to the difference between
the P48 Mould R filter and the P60 SDSS r filter. 
The ``S'' ticks on the top axis denote spectroscopic observation epochs. The solid curves are SALT2 best-fit light curves in corresponding
filters. The dashed-dotted curves are the IR template from \citet{sga+15}.
The inset zooms into the very early phases of the PTF R-band light curve. The dashed curve in the inset shows the best $t^2$ law
fit to the early light curve. 
\label{fig:lightCurve}}
\end{figure}

\begin{deluxetable*}{ccccccc}
\centering
\tablecolumns{4}
\tablewidth{0pt}
\tablecaption{{\it Swift} Observations\label{tab:swift}}
\tablehead{
    \colhead{Obs. Date} & \multicolumn{2}{c}{UVOT/$uvm2$\tablenotemark{1}} & \multicolumn{2}{c}{UVOT/$uvw2$\tablenotemark{1}} &
    \multicolumn{2}{c}{XRT\tablenotemark{2}} \\
    \colhead{}   &  \colhead{exp. time (s)}  &  \colhead{mag (AB)}  &  \colhead{exp. time (s)}  &  \colhead{mag (AB)}  &  \colhead{exp. time (s)}  &   \colhead{counts ($\textrm{cnts}\,\textrm{s}^{-1}$)}
}
\startdata
May 17.4  &  $1386$  &  $21.02\pm0.08$  &  $1540$  &  $20.59\pm0.08$  &  $2971$  &   $<3.6\times10^{-3}$  \\
May 22.6  &  $1154$  &  $21.25\pm0.10$  &  $1193$  &  $20.92\pm0.09$  &  $2382$  &   $<8.0\times10^{-3}$  \\
May 25.9  &  $1298$  &  $21.83\pm0.12$  &  $1274$  &  $21.39\pm0.10$  &  $2625$  &  $<4.3\times10^{-3}$  \\
June 02.7 &  $646$    &  $<22.20$             &  $604$    &  $<22.36$             &  $1264$  &  $<9.0\times10^{-3}$  \\
June 10.5 &  $462$    &  $<21.98$             &  $464$    &  $<22.18$             &  $928$    &  $<1.2\times10^{-2}$  \\
June 13.4 &  $612$    &  $<22.37$             &  $674$    &  $<22.46$             &  $1326$   &  $<8.7\times10^{-3}$  \\
\enddata
\tablenotetext{1}{We used the \texttt{HEASoft} package to perform aperture photometry on the UVOT images. The photometry is corrected for 
coincident loss and with the PSF growth curve and calibrated with the latest calibration \citep{BLH2011}. In order to remove host galaxy contamination
in the photometric measurements, we acquired post-SN reference frames. In cases of non-detection, we estimated 3-$\sigma$ upper limits.}
\tablenotetext{2}{We used the \texttt{XImage} software to analyze the XRT data. In cases of non-detection, we estimate upper limits with a $99.7\%$ 
confidence level.}
\end{deluxetable*}

\begin{figure}
\centering
\includegraphics[width=0.5\textwidth]{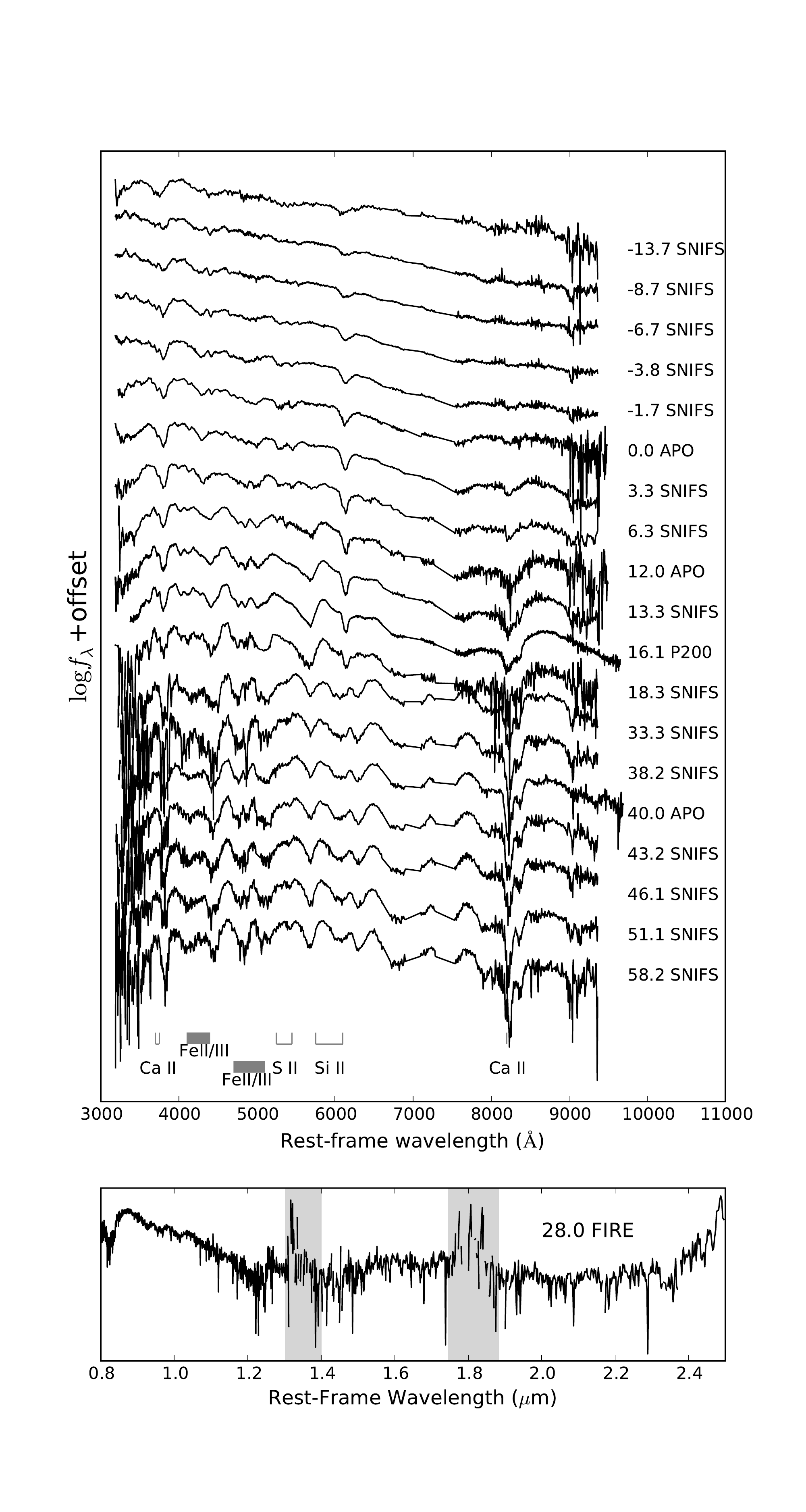}
\caption{Optical and near-IR spectral evolution of iPTF13asv. Ticks at the bottom denote the main spectral features. 
The phases and telescopes/instruments are labeled to the right of corresponding spectra. The long-slit spectra taken by APO/DIS, P200/DBSP and
Magellan/FIRE are extracted through the usual procedures in IRAF and/or IDL and calibrated using observations of spectroscopic flux standard stars. Data reduction 
of SNIFS is outlined in \citet{AAB2006}. However, due to bad weather, flux calibration of SNIFS spectra was not complete. Therefore, we interpolate
the multi-band light curves and ``warp'' the spectra with low-order polynomials to match photometric data. 
\label{fig:specEvolution}}
\end{figure}

Spectroscopic observations were undertaken with the SN Integral Field Spectrograph (SNIFS; \citealt{SNIFS}) on the $2.2\,$m telescope 
of the University of Hawaii, the Dual Imaging Spectrograph (DIS) on the ARC $3.5$\,m telescope at Apache Point Observatory (APO), the Double Spectrograph 
(DBSP; \citealt{DBSP}) on the 200-inch Hale telescope (P200) at Palomar Observatory, and the Folded-port InfraRed Echellette (FIRE) on the 
Magellan Baade Telescope at Las Campanas Observatory. The spectral sequence is presented in Figure \ref{fig:specEvolution}.

The light curves and spectra are made publicly available via WISeREP\footnote{WISeREP is available at \url{http://www.weizmann.ac.il/astrophysics/wiserep/}.} \citep{YG2012}. 

\section{Analysis}
\label{sec:analysis}

\subsection{Initial Rise and Explosion Date}
\label{sec:explosion}

In order to determine the explosion date of iPTF13asv, 
we follow \citet{NSC2011} and model the early PTF R-band light curve of iPTF13asv as a freely expanding fireball where
the luminosity increases as $\propto t^2$ and the temperature remains constant. 
Restricting ourselves to the light curve within four days of discovery, 
we find a best fit (the inset in Figure \ref{fig:lightCurve}) at an explosion date of April $29.4\pm0.3$ ($95\%$ confidence interval) 
with a fitting $\chi^2=4.3$ for five degrees of freedom. The best-fit light curve is also consistent with 
the non-detection upper limit on April $30.4$. 

If we generalize the $t^2$ model with a power-law model, we obtain strong degeneracy between the 
explosion date and the power-law index over a large range of the parameter space. In fact, \citet{FSG2015}
analyzed the rise behavior of a large sample of SNe Ia with the power-law model and also found 
that the power-law indices have large uncertainties with a mean value of $2.5$. This is probably because shallow-deposited $^{56}$Ni heats
up SN photospheres. Furthermore, there could be a dark time between the SN explosion and the SN light curve powered by radioactive
decay on the diffusive timescale for the shallowest deposition of $^{56}$Ni in the ejecta \citep{p12,pm15}. Hence,
it is nontrivial to estimate the exact explosion date purely from the early light curve. 
Since the following analysis and discussion are not very sensitive to the exact explosion date, for simplicity, we adopt the explosion 
date determined by the $t^2$ model. 

\subsection{Absence of Iron in Early-phase Spectra}
\begin{figure}
\centering
\includegraphics[width=0.45\textwidth]{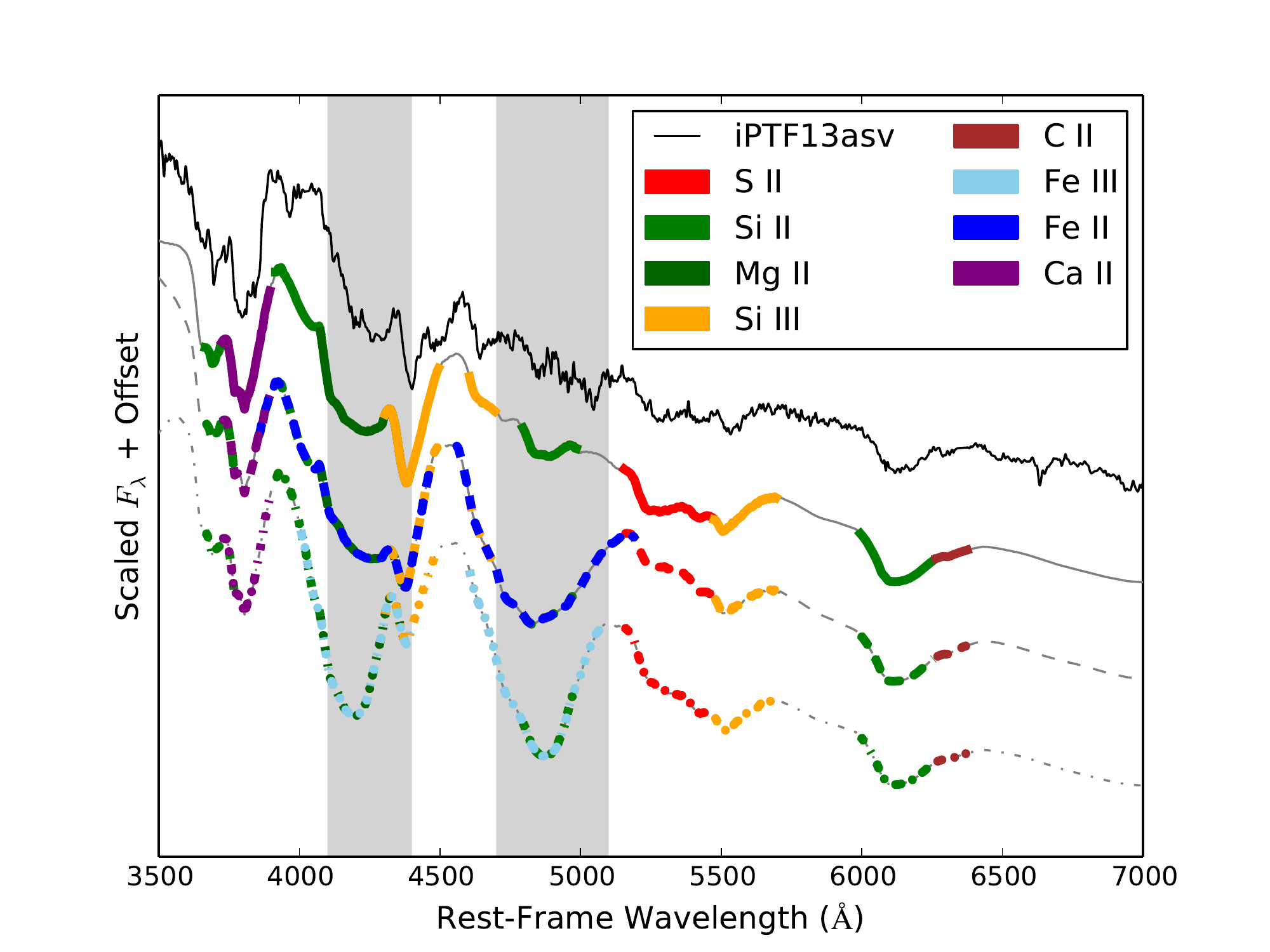}
\caption{SYN++ synthetic SN spectrum fit to iPTF13asv at -9.2 days. From top to bottom are
the observed spectrum, the synthetic spectrum without Fe (solid), the synthetic spectrum with \ion{Fe}{2} (dashed), and the synthetic
spectrum with \ion{Fe}{3} (dashed-dotted). The absorption features from different species are illustrated 
in different colors. The \ion{Fe}{2} and \ion{Fe}{3} absorption wavelength ranges are highlighted in light gray. 
\label{fig:syn++}}
\end{figure}

The most striking feature of these early-phase spectra is the absence of iron absorption. We used SYN$++$ \citep{tnm11} to perform
spectral feature identification on the spectrum taken 11 days after explosion (or equivalently, $-13.7$ days with respect to the
B-band maximum which is determined in \S\ref{sec:analysis:lightcurve}). 
As highlighted in gray in Figure \ref{fig:syn++}, the spectrum shows no signature of either \ion{Fe}{2} or \ion{Fe}{3}. 

\begin{figure*}
\centering
\includegraphics[width=0.95\textwidth]{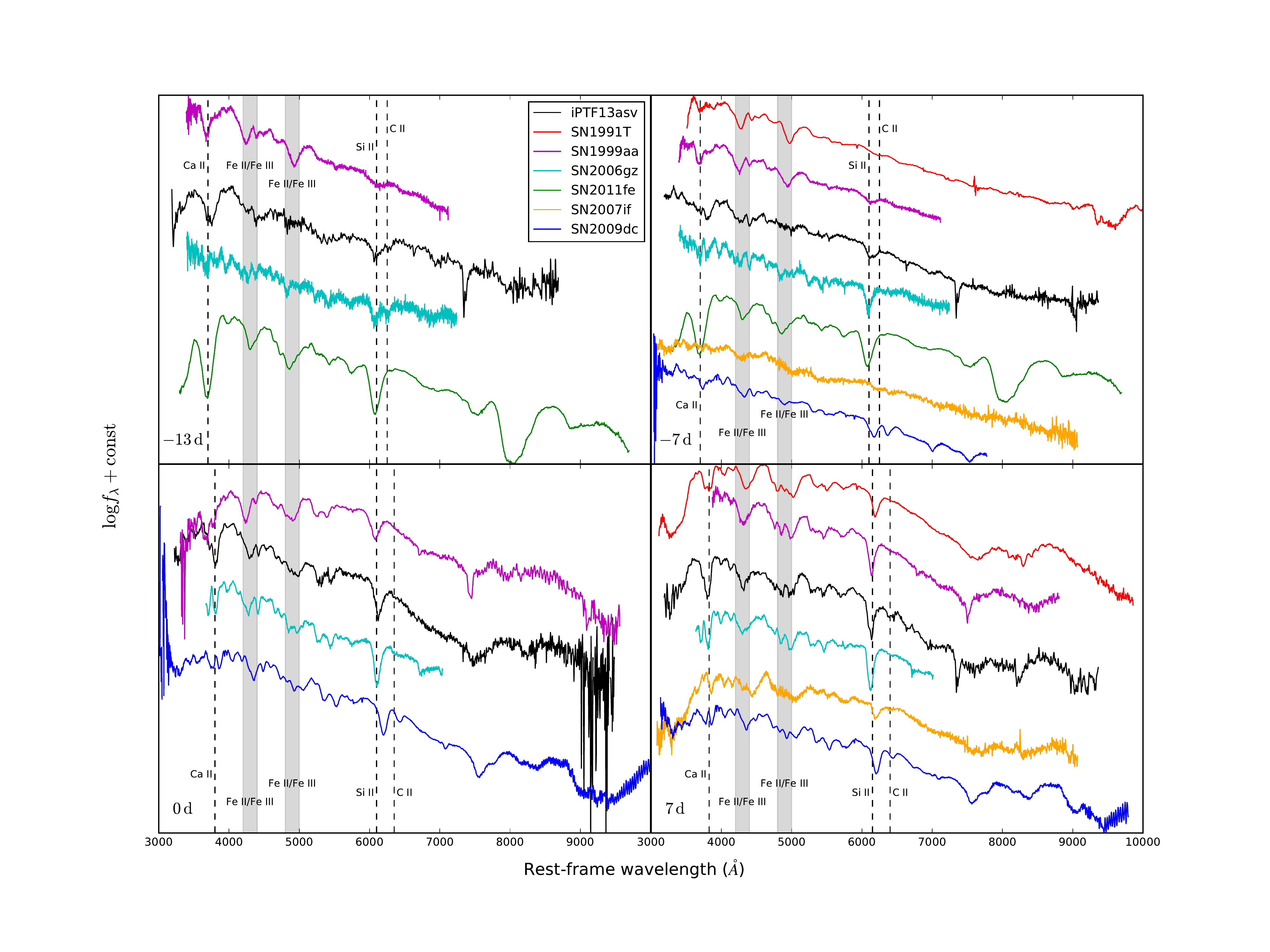}
\caption{Spectral comparison of iPTF13asv to normal SN2011fe \citep{PTA2013}; overluminous SN1991T \citep{MDT1995} and 
SN1999aa \citep{mkc+08}; and super-Chandrasekhar events SN2006gz \citep{hgp+07}, SN2007if \citep{bmk+12}, and 
SN2009dc \citep{tbc+11}. The phases of the spectra are shown at the lower left corner of each panel. 
\label{fig:specComparison}}
\end{figure*}

We also compare early-phase spectra of iPTF13asv to those of well-studied normal and overluminous Type \RNum{1}a events in the top
panels of Figure \ref{fig:specComparison}. Most spectra in comparison, including the spectrum of the super-Chandrasekhar SN2009dc 
at $-7$ days, clearly show the existence of iron absorption features. 
The exceptions are two super-Chandrasekhar events, SN2006gz and SN2007if, which have weak or no 
iron absorption.  

As a result of nucleosynthesis and mixing during SN explosions, iron commonly manifests itself as absorption features in SN spectra, either as
\ion{Fe}{2} at low effective temperatures or as \ion{Fe}{3} at higher temperatures. 
The absence of iron at early phases implies that weak mixing during the SN explosion confines synthesized iron group elements in low-velocity regions 
of the ejecta. 

The centric concentration of iron can be verified by strong UV emission at the same time, as the iron group elements are the main absorbers of photons 
below $3500$\,\AA. However, we did not trigger \textit{Swift} observations at early phases because the SN is located beyond our trigger
criteria of 100\,Mpc. In comparison to the spectral shape of SN2011fe, the spectral shape of iPTF13asv at $-7$ days (top right panel of Figure 
\ref{fig:specComparison}) indicates stronger fluxes at shorter wavelengths, hinting a strong emission in the UV. 

The weak mixing of iron in the SN explosion may have strong implications for the explosion mechanism and will be discussed in \S\ref{sec:discussion:progenitor}.
As shown in the bottom two panels of Figure \ref{fig:specComparison}, iron features appear in the iPTF13asv spectra around and after
maximum. In the next few subsections, we investigate the specifics of iPTF13asv. 

\subsection{Light Curves}
\label{sec:analysis:lightcurve}
In order to determine the light curve shape parameters of iPTF13asv, we use the SALT2 software \citep{GAB2007} to fit its optical light curve (see
Figure \ref{fig:lightCurve} for the SALT2 best-fit light curves). 
The best-fit light curve gives a rest-frame B-band peak magnitude 
$m_B=16.28\pm0.03$ on May $18.12\pm0.09$. We set this B-band peak date as $t=0$ in the rest of this paper. The fit also gives a color 
term $c=-0.16\pm0.02$, and two shape parameters $x_0=0.0055\pm0.0001$ and $x_1=0.37\pm0.09$. Based on the fitted SALT2 light curve,
we derive a color $(B-V)_0=-0.14\pm0.03$ at the B-band maximum. We also obtain $\Delta m_{15}=1.03\pm0.01$ from $x_1$ by using the relation 
in \citet{GAB2007}. 

\begin{figure}
\centering
\includegraphics[width=0.45\textwidth]{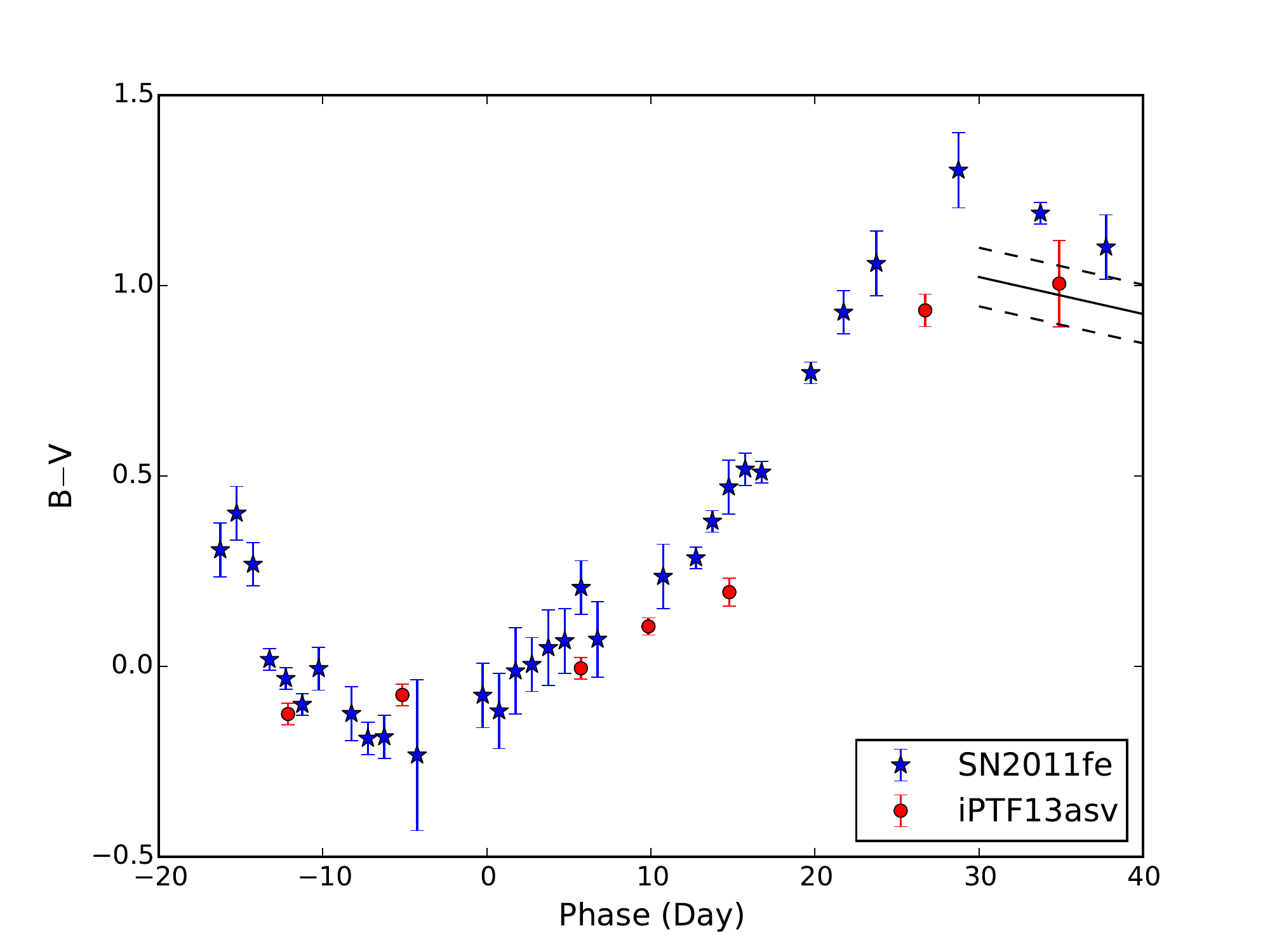}
\caption{B$-$V color evolution of SN2011fe and iPTF13asv after correction for Galactic extinction. The solid line
shows the Lira relation from \citet{bsp+14} and the dashed lines corresponding to its $0.06$\,mag scattering. 
\label{fig:BV}}
\end{figure}

The local extinction in the host galaxy of iPTF13asv is probably minor for several reasons. First, Figure \ref{fig:BV} compares the B$-$V colors
between iPTF13asv ($\Delta m_{15}=1.03\,\rm{mag}$) and SN2011fe ($\Delta m_{15}=1.10\,\rm{mag}$). 
Since SN2011fe is unreddened by its host \citep{NSC2011,vst+12}, similar colors around maximum suggest that
iPTF13asv also has little local extinction. After the maximum, iPTF13asv has a slightly blue color compared to SN2011fe, 
probably due to the different stretch of these two events \citep{ng08}, until they join the Lira relation after $+30$ days.
Second, the intrinsic color $B-V=0.95$\,mag of iPTF13asv at $+35$ days is consistent with the latest calibration of the Lira relation \citep{bsp+14}.
Third, the absence of \ion{Na}{1}\,D absorption in the low-resolution optical spectra also implies weak extinction in the host galaxy. Given a typical velocity
dispersion of $10\,\rm{km}\,\rm{s}^{-1}$ for a dwarf galaxy (see \S\ref{sec:result:host} for a discussion of the iPTF13asv host galaxy; \citealt{wmo+07}), we derive 
from the highest signal-to-noise ratio spectrum that the equivalent width for each of the \ion{Na}{1} D lines is less than $0.2\,$\AA (5-$\sigma$). 
Using the empirical relation in \citet{PPB2012}, we find that the extinction $E(B-V)<0.06$. Therefore, in what follows, we neglect the local extinction correction. 

\begin{figure*}
\centering
\includegraphics[width=0.95\textwidth]{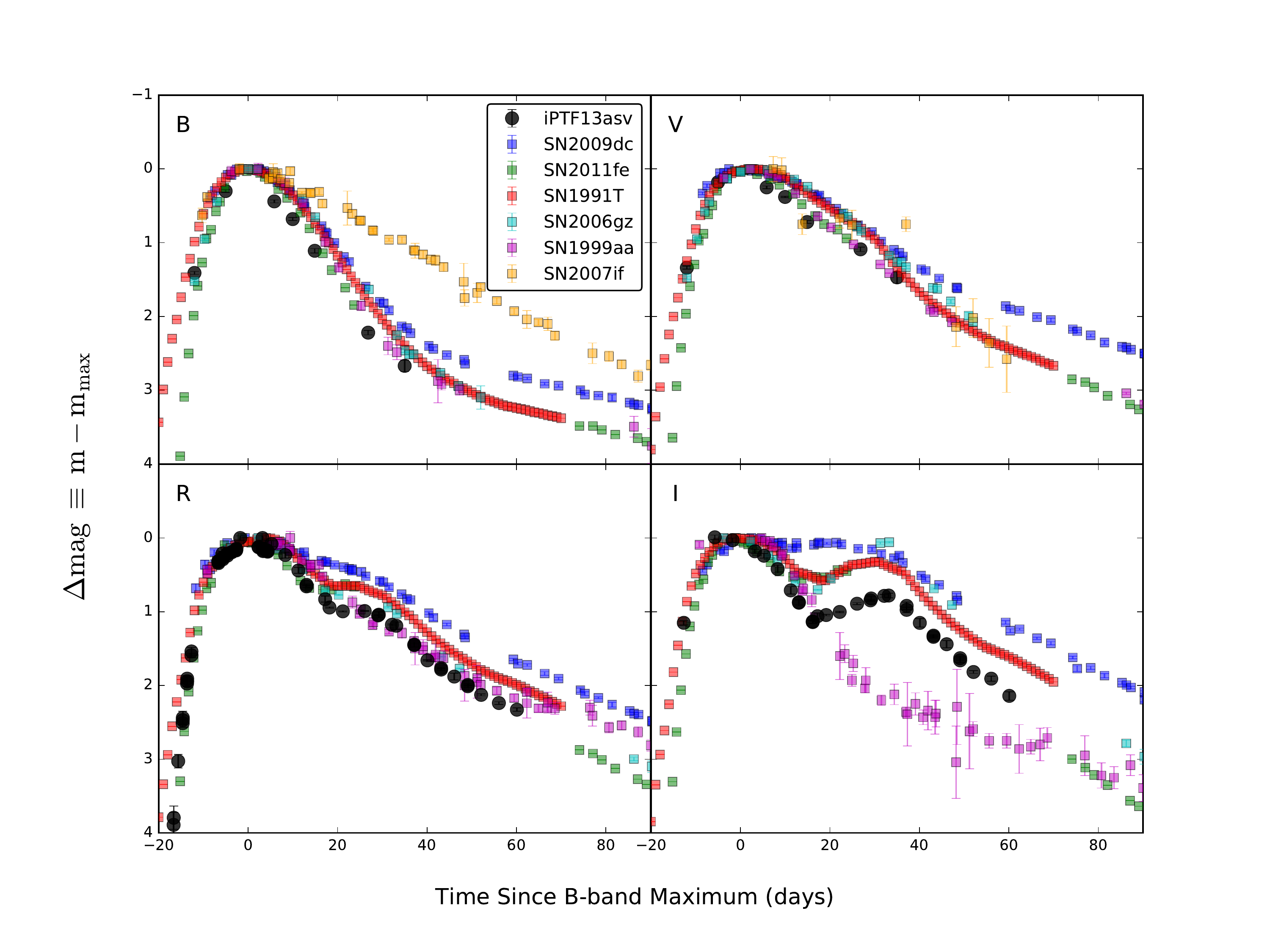}
\caption{Light curve comparison of iPTF13asv to normal SN2011fe \citep{PTA2013}; overluminous SN1991T (Nugent template) and
SN1999aa \citep{khl+00}; and super-Chandrasekhar events SN2006gz \citep{hgp+07}, SN2007if \citep{saa+10}, and 
SN2009dc \citep{sgl+11}. The legend colors are the same as in Figure \ref{fig:specComparison}. 
\label{fig:lcComparison}}
\end{figure*}

After correction for Galactic extinction we derive an absolute peak magnitude of iPTF13asv in its rest-frame B-band to be 
$-19.84\pm0.06$. This is about $0.5$\,mag brighter than normal SNe Ia at peak. 

The $k$-correction in the UV and optical wavelengths is negligible. Synthetic photometry using both the Nugent SN \RNum{1}a template \citep{nkp02}
and the observed \textit{HST} UV spectra of SN2011fe \citep{MSH2014}. 
shows that the $k$-correction is less than $0.1$\,mag in the optical and $0.2$\,mag in the \textit{Swift}/UVOT UV filters.

In Figure \ref{fig:lcComparison}, the optical light curves of iPTF13asv are compared to those of well-studied SNe, including normal SN2011fe; 
overluminous SN1999aa and SN1991T; and super-Chandrasekhar SN2006gz, SN2007if, and SN2009dc. All the light curves have been offset to match their
peak magnitudes and to the epoch of the B-band maxima. Figure \ref{fig:lcComparison} illustrates that (1) the light curve width of iPTF13asv is similar to
those of normal events and narrower than those of overluminous and super-Chandrasekhar events, except for SN2006gz, (2) iPTF13asv shows an isolated 
secondary maximum in the I-band whose strength is weaker than those observed in SN2011fe and SN1991T, and (3) iPTF13asv matches well 
to the super-Chandrasekhar SN2006gz in the B- and V-band light curves, but SN2006gz has a much stronger near-IR secondary peak. 

\begin{figure}
\centering
\includegraphics[width=0.45\textwidth]{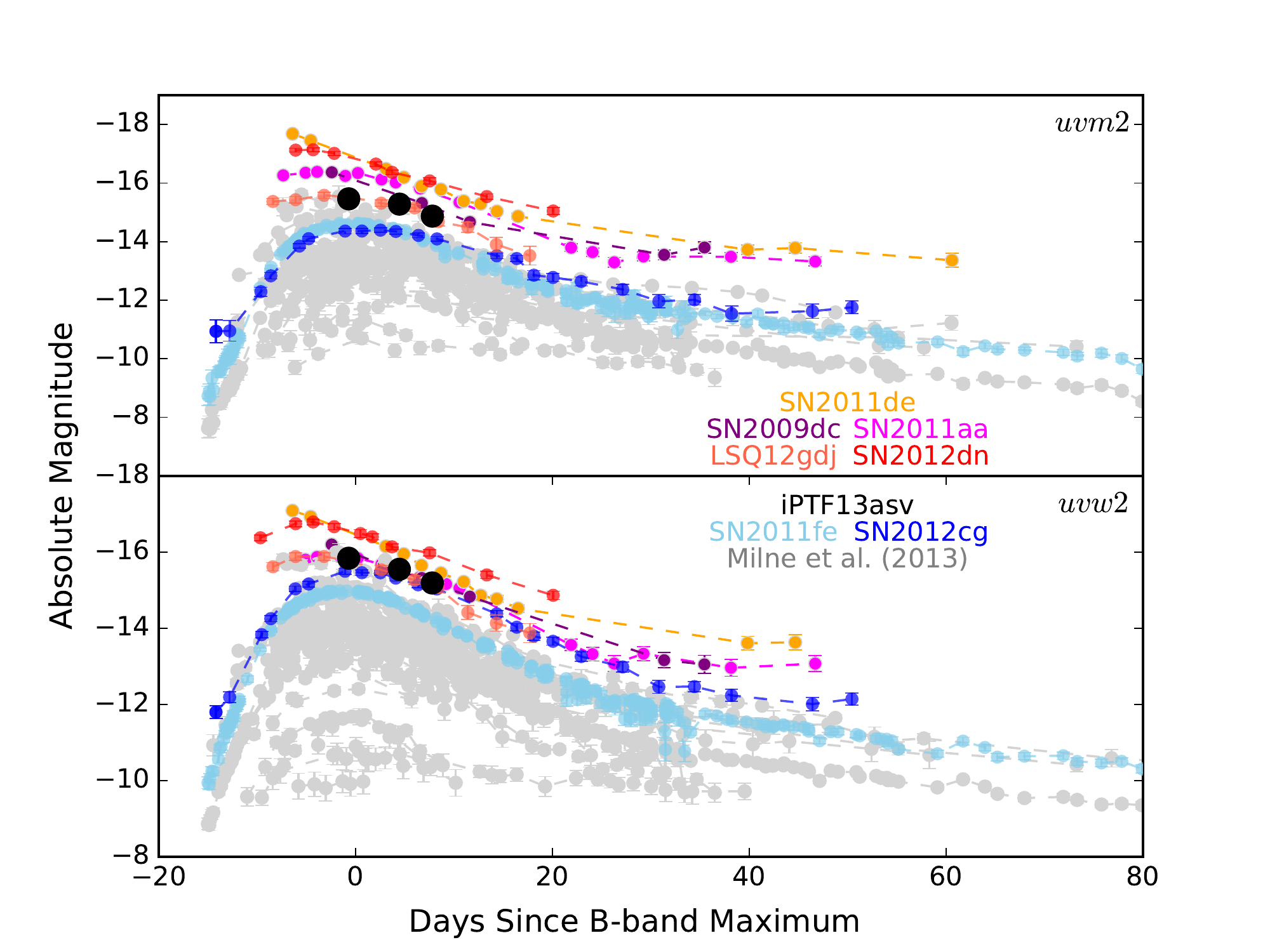}
\caption{UV light curve comparison between iPTF13asv and other SNe \RNum{1}a.
A sample of SNe \RNum{1}a from \citet{MBR2013} is shown in gray. Highlighted by different colors are light curves of iPTF13asv (black); 
SN2011fe (sky blue; \citealt{BDd2012}); SN2012cg (blue); super-Chandrasekhar events SN2009dc, SN2011aa, and
LSQ12gdj (purple; \citealt{bks+14}); and the most UV-luminous event SN2011de (orange; \citealt{Brown2014}). 
All magnitudes are in the AB system.
\label{fig:swiftIaSample}}
\end{figure}

Figure \ref{fig:swiftIaSample} compares iPTF13asv in the \textit{Swift}/UVOT $uvm2$ and $uvw2$ filters to a large sample of
both normal, overluminous, and super-Chandrasekhar SNe Ia observed by \textit{Swift} \citep{MBR2013,Brown2014}. 
While the $uvw2$ filter has a non-negligible leakage in long wavelengths, the $uvm2$ filter does not have a significant leakage
and therefore provides the best available measurements of the UV flux. 
The figure shows that, like super-Chandrasekhar events, iPTF13asv is more luminous in the UV than the majority of normal events. 
Furthermore, \citet{MBR2013} divided SNe Ia into different subclasses based on their \textit{Swift}/UVOT colors. We 
cannot make a direct comparison here because only $uvm2$ and $uvw2$ data are available for iPTF13asv. An indirect comparison is that
iPTF13asv is brighter than SN2011fe by half a magnitude in the optical and by $\simeq0.7$\,mag in the UV. Since SN2011fe belongs to 
the NUV-blue subclass \citep{BDd2012}, iPTF13asv probably also belongs to the same subclass. 

We also compare the IR light curves of iPTF13asv in the J- and H-band to the most recent light curve template for normal Type \RNum{1}a
events (Figure \ref{fig:lightCurve}; \citealt{sga+15}) and find that the sparsely sampled light curves of iPTF13asv roughly follow the template. 
The peak magnitudes of iPTF13asv is $M_J=-18.83\pm0.08$ and $M_H=-18.16\pm0.19$, compared to the median peak magnitudes of
$M_J=-18.39$ and $M_H=-18.36$ with rms scatters of $\sigma_J=0.116\pm0.027$ and $\sigma_H=0.085\pm0.16$ for normal Type \RNum{1}a
SNe \citep{blw+12}. The secondary maximum of iPTF13asv is clearly seen in J-band and H-band light curves as well as the I-band light curve, 
indicating concentration of iron group elements in the central region of the ejecta \citep{Kasen2006}. This is in accordance with the absence of
iron in the outer ejecta.

\subsection{Spectra}
\subsubsection{Spectral Cross-matching}
We use the latest versions of both SN Identification (SNID; \citealt{bt07}) and SuperFit \citep{HSP2005} to spectroscopically classify iPTF13asv. 
Before $-5$ days, not surprisingly, neither tools find a good match for iPTF13asv spectra partly because they do not have many early-phase SNe
in their templates and partly because the iron absorption is absent in the early-phase spectra of iPTF13asv. 
Around and after maximum, based on the first 5 best matches, both SNID and SuperFit find that iPTF13asv spectroscopically resembles normal 
SNe Ia (Table \ref{tab:snid}).
 
\begin{deluxetable*}{c|ccccc}
\centering
\tablecolumns{6}
\tablewidth{0pt}
\tablecaption{SNID results\label{tab:snid}}
\tablehead{
    \colhead{Phase} & \multicolumn{5}{c}{First Five Best Matches\tablenotemark{1}}
    }
\startdata
$-6.8$  &   05eu@$-5.2$ (normal)    &    03ic@$-4.1$ (normal)  &   08Z@$-4.3$  (normal) &  05na@$-1.5$ (normal) &  06cc@$-9.7$ (normal) \\
$0.0$   &   96ai@$+2.2$ (normal)      &     07F@$+3.0$ (normal)   &   94ae@$0.0$ (normal)  &  03cg@$-2.1$ (normal)  &  94ae@$+0.9$ (normal) \\
$+6.8$   &   08Z@$+7.6$ (normal)     &     05na@$+3.4$ (normal)   &   99aa@$+1.8$ (peculiar)  &  01fe@$+6.2$ (normal)  &  06cz@$-2.0$ (91T-like) \\
$+13.3$  &  08Z@$+12.3$ (normal)    &    07ca@$+13.3$ (normal)  &  07bj@$+12.0$ (normal)  &  03fa@$+13.4$ (91T-like)  &  03kf@$+13.4$ (normal)
\enddata
\tablenotetext{1}{The format in this column is name@phase (subclass).}
\end{deluxetable*}

\subsubsection{Spectral comparison to well-studied SNe}
\label{sec:snid}
Figure \ref{fig:specComparison} compares iPTF13asv spectra to those of well-studied SNe at different epochs: normal SN2011fe \citep{PTA2013}; 
overluminous SN1991T \citep{frm+92} and SN1999aa \citep{GFG2004,mkc+08}; and super-Chandrasekhar SN2006gz \citep{hgp+07}, SN2007if \citep{saa+10},
and SN2009dc \citep{tbc+11}. At $-13$ days (top left panel of Figure \ref{fig:specComparison}), although both SN2006gz and iPTF13asv have weak or no 
absorption from iron, iPTF13asv shows strong \ion{Ca}{2}
H and K absorption but SN2006gz does not. 
In comparison, both SN1999aa
and SN2011fe at similar phases have both strong \ion{Ca}{2} and iron absorptions. 
In addition, the absorption of \ion{C}{2} is apparently weaker in iPTF13asv than in SN2006gz.

At a week before maximum (top right panel of Figure \ref{fig:specComparison}), except for the prominent absence of iron in iPTF13asv, 
the overall spectral features of iPTF13asv are similar to those of normal SNe. Unlike the near-absence of \ion{Ca}{2} and \ion{Si}{2} lines in
SN1991T, iPTF13asv shows apparent \ion{Ca}{2} and \ion{Si}{2} absorption, the strengths of which are weaker than those seen SN1991T. 
Besides this, its \ion{C}{2} feature becomes weaker.

Around maximum (lower left panel of Figure \ref{fig:specComparison}), we find good spectral matches between iPTF13asv and SN2011fe.
At this epoch, SN1991T is also becoming similar to normal SNe Ia. The strength of \ion{Si}{2} absorption in iPTF13asv is between 
the weak absorption in SN1991T and the strong one in SN2011fe.

One week after maximum (lower right panel of Figure \ref{fig:specComparison}), the spectrum of iPTF13asv is very similar to those of normal
events, but with strong \ion{C}{2} absorption.

\subsubsection{\ion{Si}{2} velocities}
\label{sec:velocity}
We further measure the expansion velocity evolution of iPTF13asv by fitting a Gaussian kernel to the \ion{Si}{2}\,6355 line in each spectrum. The 
continuum is modeled by a linear regression to regions at both sides of the line. Then we fit a linear model to the velocity measurements between
$-10$ and $+10$ days and estimate a velocity of $(1.0\pm0.1)\times10^{4}\,\rm{km}\,\rm{s}^{-1}$ and a velocity gradient close to zero at peak. 

\begin{figure}
\centering
\includegraphics[width=0.45\textwidth]{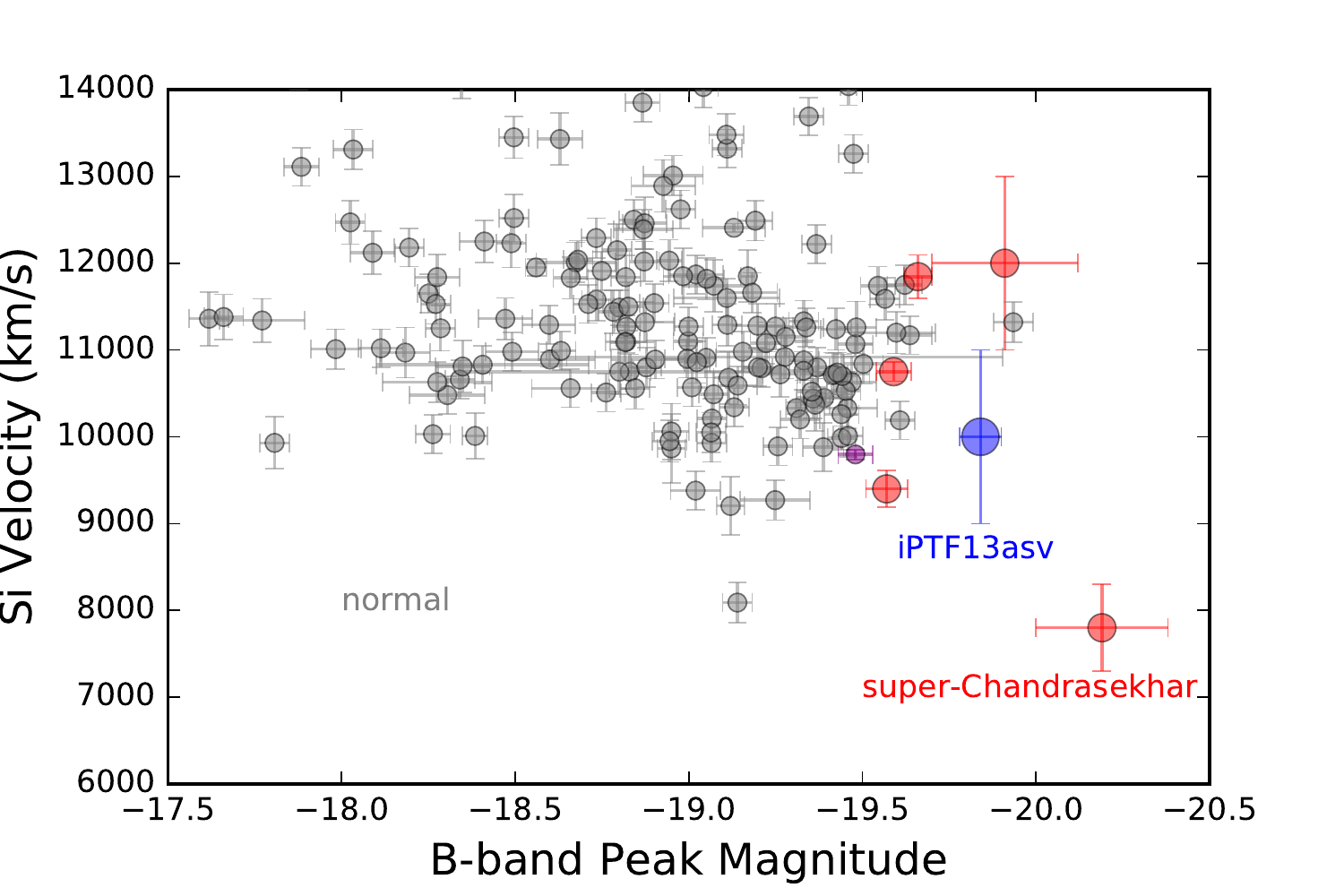}
\caption{\ion{Si}{2} velocities at maximum vs. peak magnitudes. The gray points are measurements taken from \citet{fsk11} and \citet{sra14}. 
Note that \citet{fsk11} did not correct the local extinction. The red points are taken from \citet{saa+12}.
\label{fig:velocity}}
\end{figure}

\begin{figure}
\centering
\includegraphics[width=0.45\textwidth]{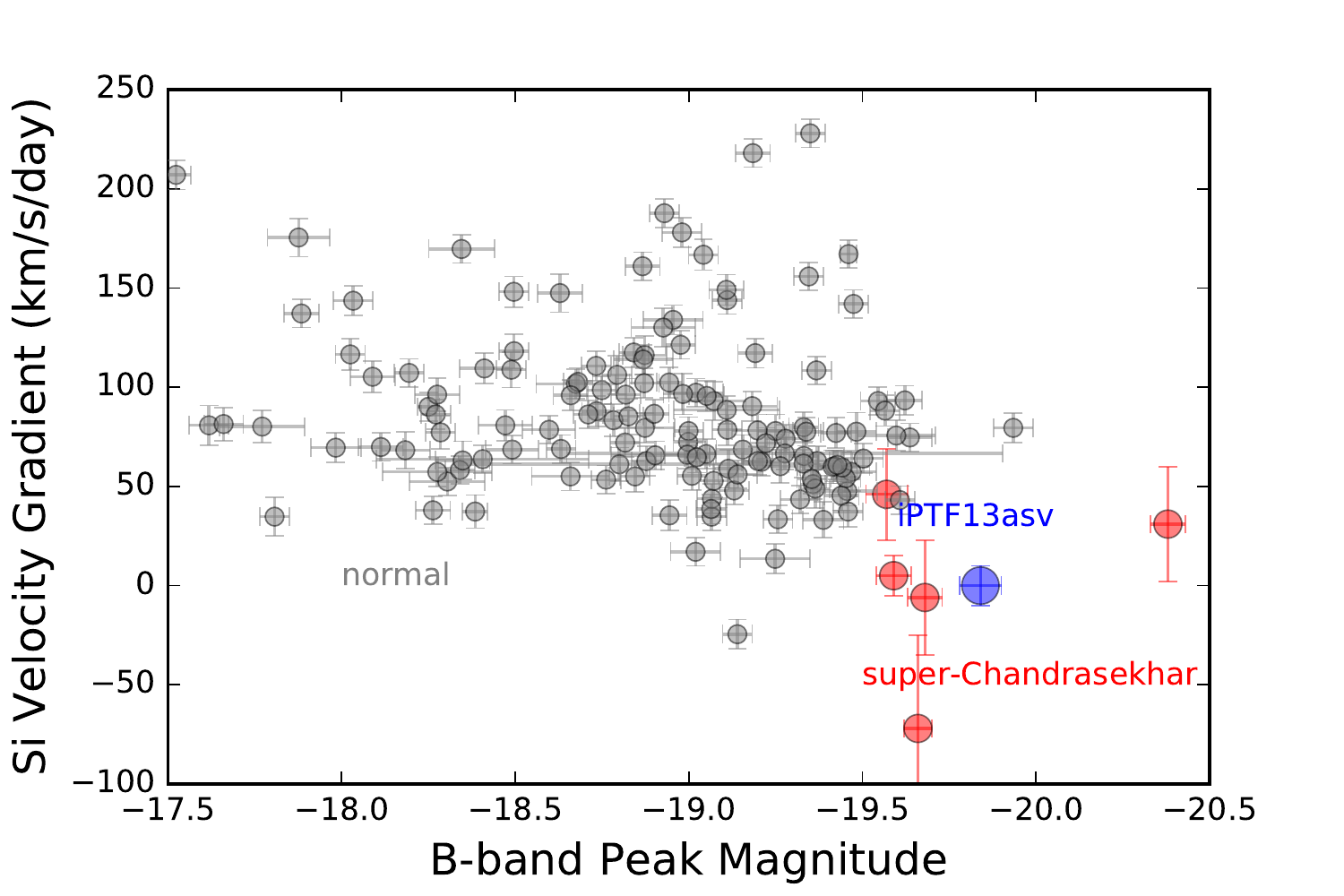}
\caption{\ion{Si}{2} velocity gradients at maximum vs. peak magnitude. The gray points are measurements taken from \citet{fsk11}. 
Note that \citet{fsk11} did not correct the local extinction. The red points are taken from \citet{saa+12}.
\label{fig:velocityGradient}}
\end{figure}

Figure \ref{fig:velocity} shows that \ion{Si}{2} velocities at maximum versus peak
magnitudes. As can be seen in the figure, iPTF13asv has a \ion{Si}{2} velocity lower than the majority of normal SNe Ia and similar
to super-Chandrasekhar events. Figure \ref{fig:velocityGradient} compares \ion{Si}{2} velocity gradients at maximum versus peak magnitudes. 
Again, like super-Chandrasekhar events, iPTF13asv has a velocity gradient close to zero, lower than the majority of normal events.

\subsubsection{Carbon signatures}
\label{sec:carbon}

\begin{figure}
\centering
\includegraphics[width=0.5\textwidth]{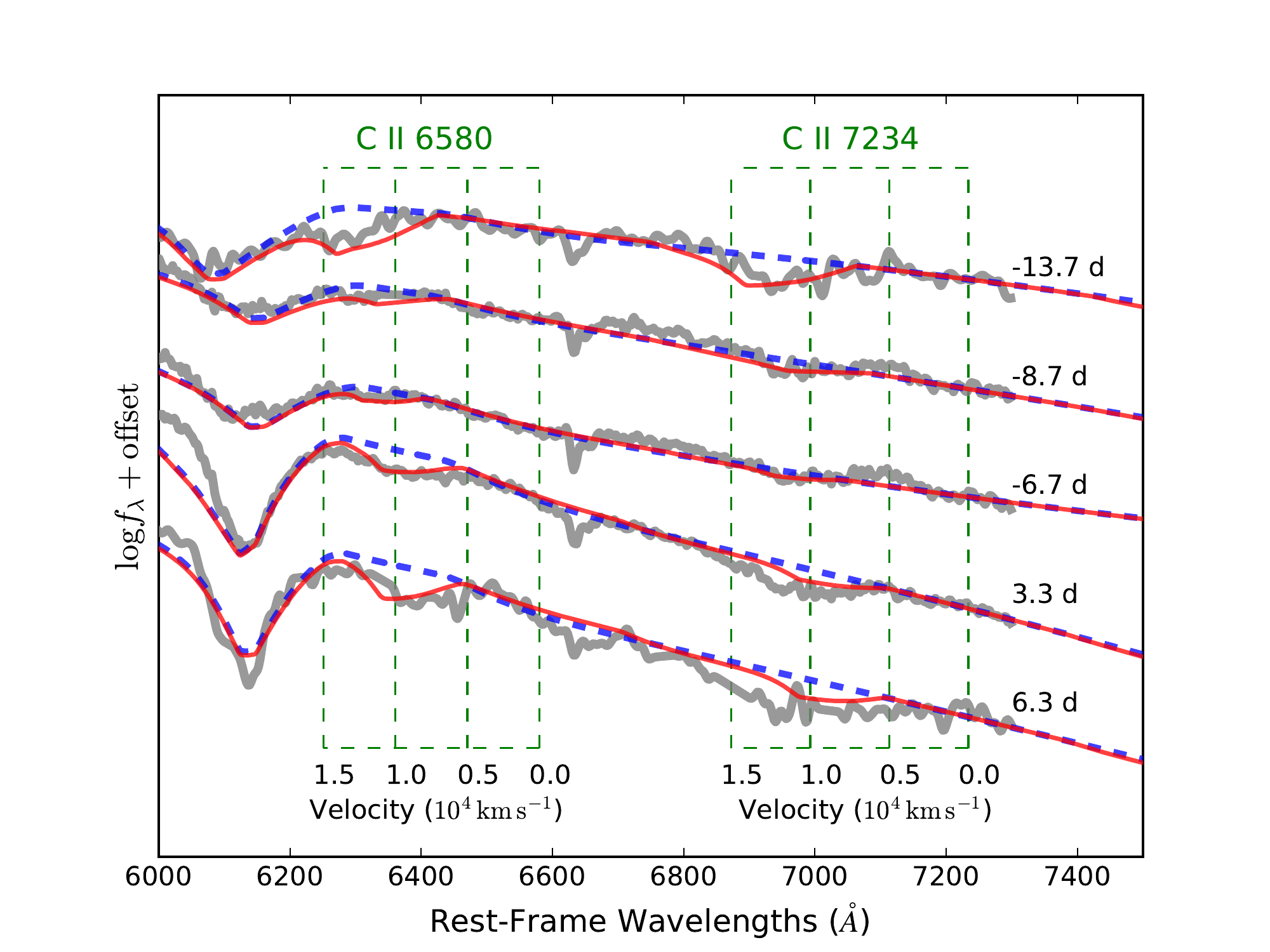}
\caption{Carbon features of iPTF13asv at different phases. The numbers to the right of each spectrum indicate the phases in days. 
The observed spectra are shown in gray. The SYN$++$ spectra without \ion{C}{2} are in blue and those with \ion{C}{2} are in red. 
The green dashed axes show velocities of \ion{C}{2}\,$6580$ and \ion{C}{2}\,$7234$ lines. 
\label{fig:carbon}}
\end{figure}

We also note that iPTF13asv shows weak but persistent \ion{C}{2} absorption features until at least a week after maximum. Figure \ref{fig:carbon} shows
SYN$++$ fits to iPTF13asv spectra of high signal-to-noise ratios, demonstrating the existence of \ion{C}{2}\,6580 and \ion{C}{2}\,7234 lines.
The velocities of these \ion{C}{2} lines evolve from $\simeq14,000\,\rm{km}\,\rm{s}^{-1}$ at $-13.7$ days to $\simeq11,000\,\rm{km}\,\rm{s}^{-1}$ at
$+6.3$ days. 

\begin{figure}
\centering
\includegraphics[width=0.5\textwidth]{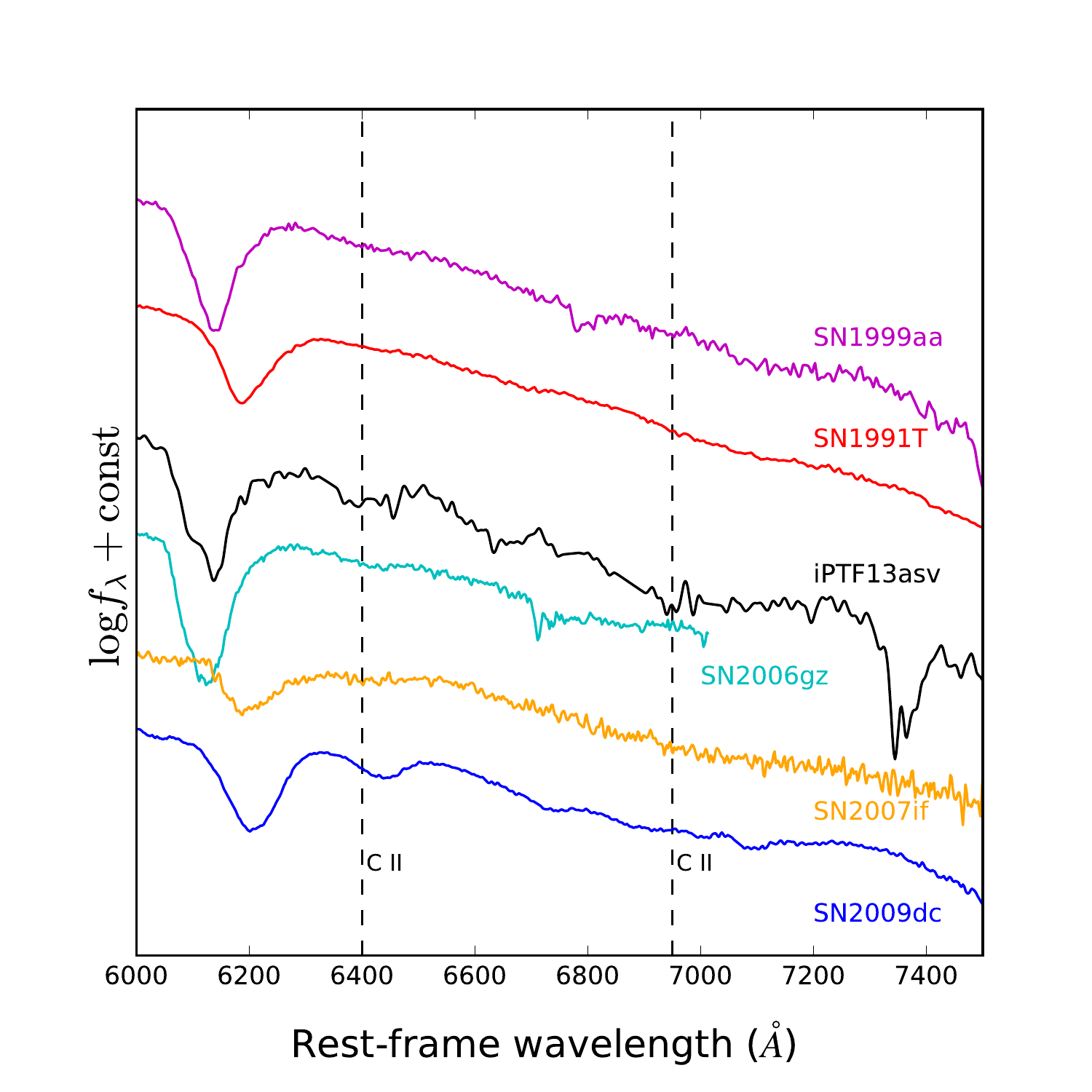}
\caption{Comparison of carbon features among iPTF13asv, SN1991T, SN1999aa, SN2006gz, SN2007if, and SN2009dc at one week after maximum.
\label{fig:carbonComparison}}
\end{figure}

About 30\% of normal SNe are estimated to reveal the \ion{C}{2}\,6580 and \ion{C}{2}\,7234 absorption notches in early phases \citep{taa+11,ptf+11,sf12}.
These \ion{C}{2} features usually disappear before maximum. In contrast, some super-Chandrasekhar events show strong and persistent \ion{C}{2}
features even after maximum. Figure \ref{fig:carbonComparison} compares the spectra of iPTF13asv at one week after maximum to those of well-studied
SNe at similar phases. As can be seen, neither SN1991T nor SN1999aa has the carbon feature at this phase; the carbon signature of iPTF13asv is not as strong as those seen in the super-Chandrasekhar SN2009dc. 

\subsection{Host Galaxy}
\label{sec:result:host}
After iPTF13asv faded away, we obtained a low signal-to-noise ratio spectrum of its apparent host galaxy SDSS\,J162254.02+185733.8. 
The spectrum only shows H$\alpha$ emission at the redshift of iPTF13asv. We fit a Gaussian profile to the H$\alpha$ line and
measure a luminosity of $3\times10^{38}\,{\rm ergs\,s}^{-1}$. We adopt the empirical relation between H$\alpha$ luminosity and 
star formation rate \citep{Kennicutt1998} and obtain a star formation rate of $2\times10^{-3}\,\sm\,{\rm yr}^{-1}$ for the host galaxy. 

\begin{figure}
\centering
\includegraphics[width=0.45\textwidth]{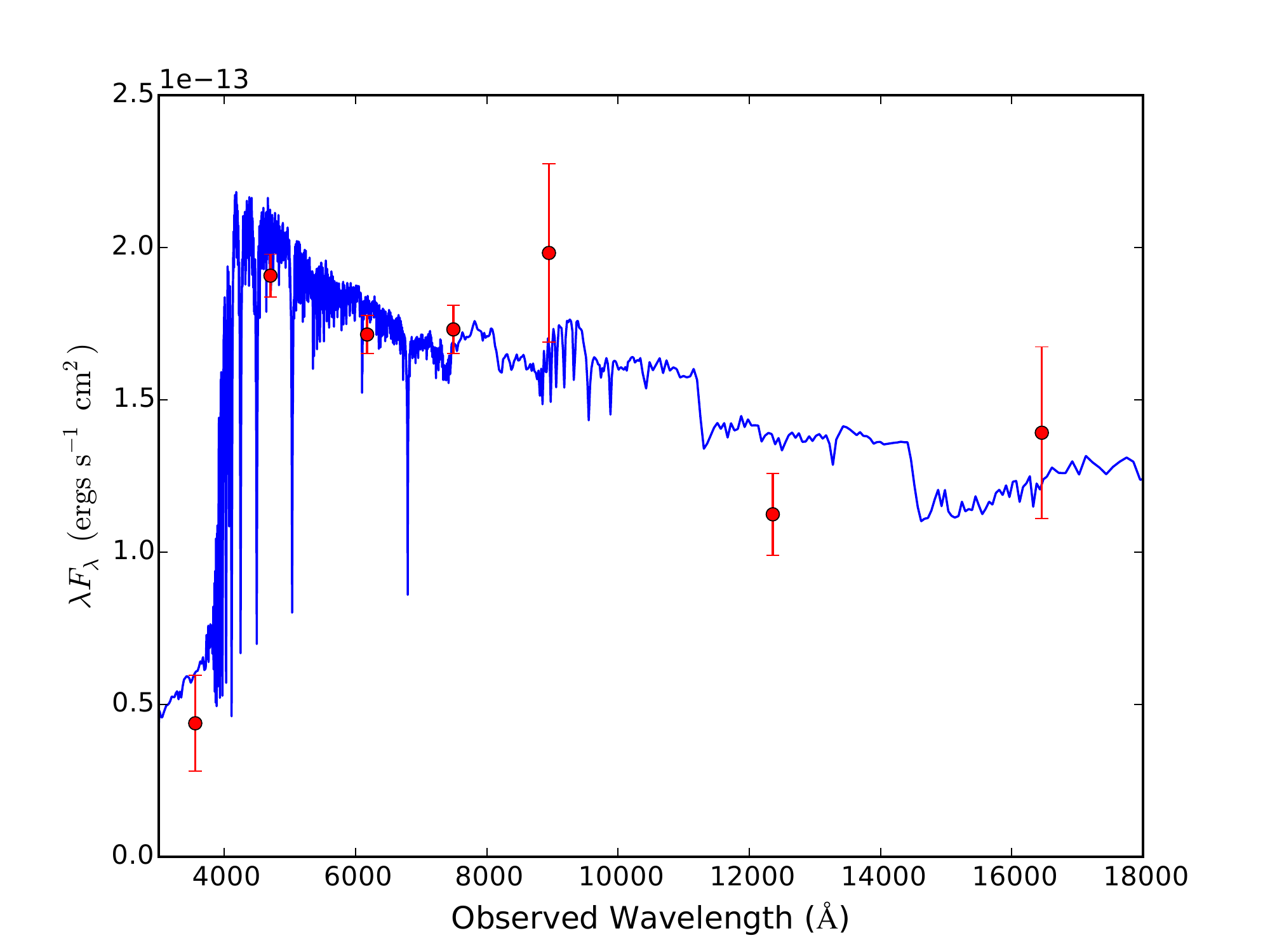}
\caption{SED fit of the host galaxy. The data points (red) are SDSS model magnitudes in optical and aperture-photometric measurements 
in the near-IR RATIR reference images. The blue spectrum is the best fit from FAST. 
\label{fig:sedfit}}
\end{figure}

Next, we construct the spectral energy distribution (SED) of the host galaxy with optical photometry from SDSS and near-IR photometry 
measured on the SN reference images.  
The SED is then modeled with a galaxy synthesis code called the Fitting and Assessment 
of Synthetic Templates (\citealt{kvl+09}) assuming an exponentially decaying star formation history and a solar metallicity. 
The best-fit model gives a galaxy age of $10^{8.6}\,$years and a stellar mass 
$\log_{10}\left(M_{\rm stellar}/\sm\right)=7.85^{+0.5}_{-0.4}$ with a reduced $\chi^2=1.6$ (Figure~\ref{fig:sedfit}). 
The best-fit model also shows no ongoing star forming activity. Because SED fitting models are usually insensitive to very low
star-forming rates, the best-fit model is consistent with the low star formation
rate derived from the H$\alpha$ flux. The derived star formation rate and the stellar mass of the iPTF13asv host galaxy 
follow the empirical relation between stellar mass and star formation rate \citep{fhg+12}. 

Since the host galaxy spectrum does not show [\ion{N}{2}] lines, we estimate an upper limit of $\log\left(\right.$[\ion{N}{2}\,6548/H$\alpha$]$\left.\right)<-0.87$. 
Using \citet{dtt02}, we derived a metallicity upper limit of $12+\log\left(\rm{O}/\rm{H}\right) < 8.3$. In fact, 
using the mass-metallicity relation \citep{fhg+12}, we estimate a gas-phase metallicity of $12+\log({\rm O/H})\sim8$ for the host galaxy. 
Compared to the host galaxy samples of SNe Ia in \citet{PSM2014} and \citet{wdg+16}, SDSS\,J162254.02+185733.8 is one of the least massive and most metal-poor
galaxies that host SNe Ia. 

\section{Bolometric Light curve And Ejecta Mass}
\label{sec:mass}

\subsection{Bolometric Light curve}
Given the wavelength coverage of the iPTF13asv spectra, we first construct a pseudo-bolometric light curve between 3500 and 9700\,\AA. In order
to calibrate the absolute fluxes of these spectra, we use interpolated optical light curves to ``warp'' the spectra. Then the spectra are integrated to
derive the pseudo-bolometric light curve. 

Due to the sparsely sampled UV and IR light curves, it is difficult to estimate the UV and IR radiation at different phases. Therefore we calculate optical-to-bolometric 
correction factors with a spectral template \citep{hch+07}. In this calculation, we find that the UV correction reaches about 25\% before the B-band 
maximum and quickly drops to less than 5\% around and after the B-band maximum. 
Given the inference that the SN might be UV-luminous before maximum and the observational fact that the SN is among the UV-bright Type \RNum{1}a
SNe around maximum, our calculated correction probably underestimates the UV radiation. Using the \textit{Swift} data around maximum, we estimate that 
this UV correction introduces a systematic uncertainty of a few percent to the bolometric luminosity. Around maximum when the SN cools down, the
UV contribution to the bolometric luminosity becomes even less important. 

In the IR, the correction above 9700\,\AA\ is below 10\% around the B-band maximum, and then reaches a maximum of 24\% around the secondary maximum 
in the near-IR. At the epochs with IR data, we find that the calculated correction is consistent with the IR measurements. 

\begin{figure}
\centering
\includegraphics[width=0.45\textwidth]{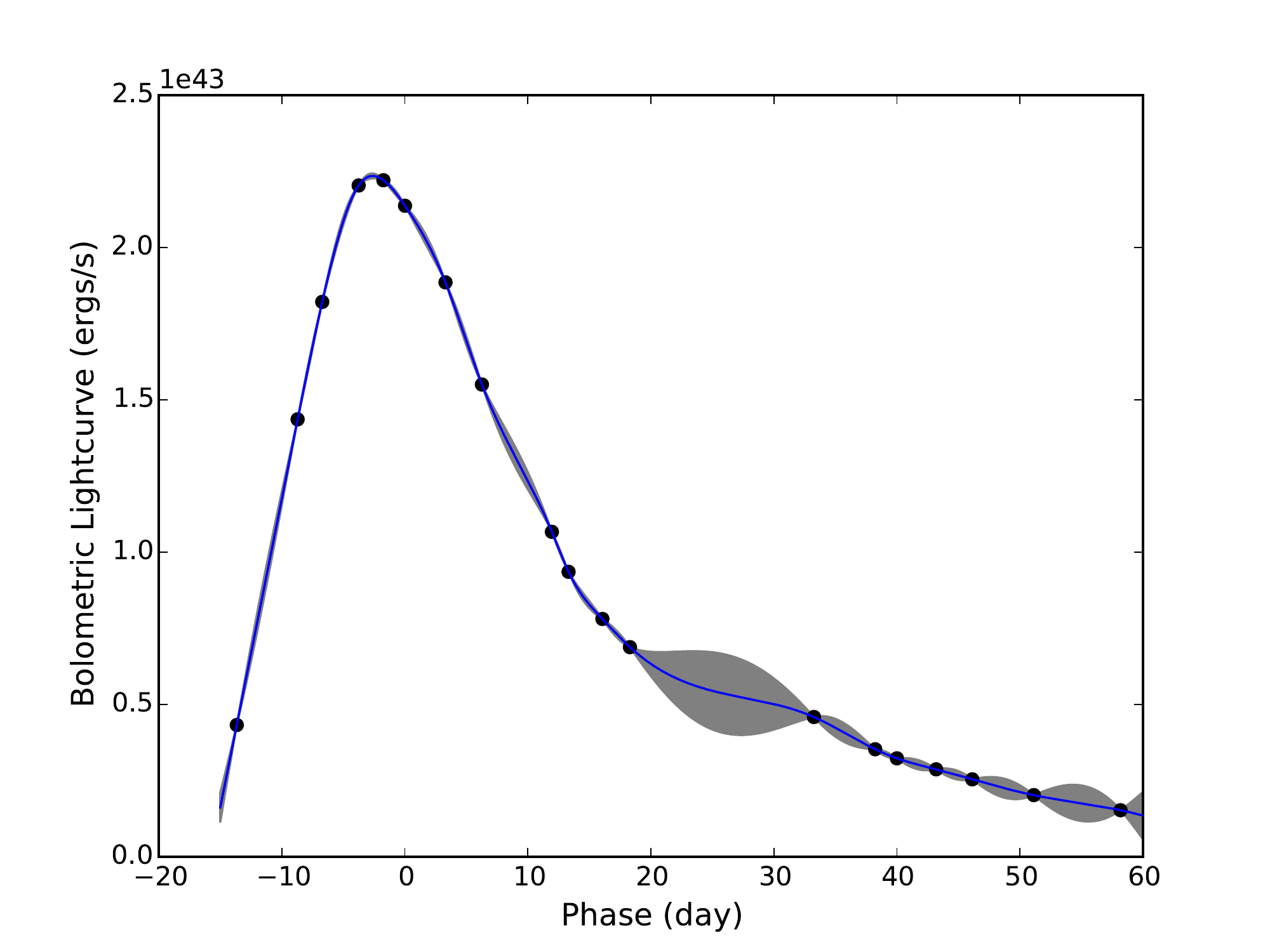}
\caption{Bolometric light curve. The measured bolometric luminosities at different phases are in black circles. The blue curve is the best fit from the Gaussian 
process regression. The gray region represents the 1-$\sigma$ uncertainty of the regression curve. 
\label{fig:L_bol}}
\end{figure}

The final bolometric light curve is shown in Figure \ref{fig:L_bol}. We further employ Gaussian process regression to derive
a maximum bolometric luminosity $L_{max}=(2.2\pm0.2)\times10^{43}\,\rm{erg}\,\rm{s}^{-1}$ at $-2.6$ days.

\subsection{$^{56}$Ni Mass and Ejecta Mass}
\label{sec:mass:mass}

Next, we follow the procedure in \citet{saa+12,SAA2014} to derive the $^{56}$Ni mass and the total ejecta mass. 
First, the $^{56}$Ni mass can be estimated through the following equation
\begin{equation}
L_{max} = \alpha S(t_R)\ ,
\end{equation}
where $S(t_R)$ is the instantaneous radioactive power at the bolometric luminosity maximum. $\alpha$ is an efficiency factor of order unity, depending on
the distribution of $^{56}$Ni \citep{jbb06}. We adopt a fiducial value of $\alpha=1.3$ following \citet{saa+12}. The radioactive
power of $^{56}\rm{Ni}\rightarrow^{56}\rm{Co}\rightarrow^{56}\rm{Fe}$ is \citep{n94}
\begin{equation}
S(t_r) = \left[6.31\exp{(-t_r/8.8)} + 1.43\exp{(-t_r/111)}\right]M_{\rm{Ni}}\ ,
\end{equation}
where $S(t)$ is in units of $10^{43}\,\rm{erg}\,\rm{s}^{-1}$ and $M_{\rm{Ni}}$ is in units of $\sm$. With the measured maximum bolometric luminosity
$L_{max}=(2.2\pm0.2)\times10^{43}\,\rm{erg}\,\rm{s}^{-1}$ at $-2.6$ days, we estimate a $^{56}$Ni mass of $(0.77\pm0.07)\sm$. 

About one month after the SN maximum, the SN debris expands approximately in a homologous manner. At this time, most $^{56}$Ni atoms have decayed to
$^{56}$Co. Hence the total luminosity can be approximated by
\begin{equation}
L(t) = \left[1-\exp(-(t_0/t)^2)\right]S_\gamma(t) + S_{e^+}(t)\ ,\label{eq:lateTime}
\end{equation}
where $S_\gamma$ and $S_{e^+}$ are the decay energy of $^{56}$Co carried by $\gamma$-ray photons and positrons. At 
time $t_0$, the mean optical path of $\gamma$-ray photons becomes unity. For a given density and velocity profile,
$t_0$ reflects the column density along the line of sight. We fit equation (\ref{eq:lateTime})
to the bolometric light curve of iPTF13asv after $+20$ days and obtained $t_0=44.2\pm2.0\,\rm{days}$. 

Next, we estimate the total ejecta mass of iPTF13asv. If we assume a density profile $\rho(v)\propto\exp(-v/v_e)$ where $v_e$
is a scale velocity, then the ejecta mass can be expressed as
\begin{equation}
M_{ej}=\frac{8\pi}{\kappa_\gamma q}\left(v_et_0\right)^2\ ,
\end{equation}
where $\kappa_\gamma$ is the Compton scattering opacity for $\gamma$-ray photons . The value of $\kappa_\gamma$ is expected to lie 
in the range between $0.025$ and $0.033\,\rm{cm}^2\rm{g}^{-1}$ \citep{ssh95}. We adopt a value of $0.025\,\rm{cm}^2\rm{g}^{-1}$ 
for the optically thin regime. The form factor $q$ describes the distribution of $^{56}$Ni and thus $^{56}$Co \citep{j99}. For evenly mixed $^{56}$Ni, 
the value of $q$ is close to one-third. Taking element stratification and mixing in the interfaces into account, \citet{SAA2014} found that $q=0.45\pm0.05$.
Here, we adopt $q=0.45$ in our estimation. 

The value of $v_e$ can be obtained by conservation of energy. The total kinetic energy of the ejecta is $6M_{ej}v_e^2$. Neglecting the radiation energy, 
the total kinetic energy is equal to the difference between the nuclear energy released in the explosion and the binding energy of the exploding  
WD. The binding energy of a rotating WD with mass $M_{ej}$ and central density $\rho_c$ is given in \citet{yl05}. Here
we restrict the central density to lie between $10^7$ and $10^{10}\,\rm{g}\,\rm{cm}^{-3}$. 

If we further assume that the 
ejecta is composed of unburned CO and synthesized Si, Fe, and Ni, then the nuclear energy of the SN
explosion is formulated in \citet{mi09} as a function of mass $M_{ej}$ and mass fractions $f_{\rm{Fe}}$, $f_{\rm{Ni}}$ and $f_{\rm{Si}}$. The ratio
$\eta=f_{\rm{Ni}}/(f_{\rm{Ni}}+f_{\rm{Fe}})$ is also a function of $\rho_c$. Following \citet{SAA2014}, we adopt a Gaussian prior
\begin{equation}
\eta=0.95-0.05\rho_{c,9}\pm0.03\max(1,\rho_{c,9})\ ,
\end{equation}
where $\rho_{c,9}$ is $\rho_c$ in units of $10^9\,\rm{g}\,\rm{cm}^{-3}$. In addition, we restrict the mass fraction $f_{\rm{CO}}$ less than $10\%$. 

Based on the above assumptions, with a given set of ejecta mass $M_{ej}$, central density $\rho_c$ and the mass fractions of different elements, 
we can calculate the maximum bolometric luminosity and $t_0$ in equation (\ref{eq:lateTime}) and compare them with our measurements of iPTF13asv.
Here we perform Markov-Chain Monte-Carlo simulations for one million steps and obtain $M_{Ni}=0.81^{+0.10}_{-0.18}$ and $M_{ej}=1.44^{+0.44}_{-0.12}\sm$ at a 95\% 
confidence level. 

\subsection{Detached Shell Surrounding the SN}

In order to explain the almost constant \ion{Si}{2} velocity in super-Chandrasekhar events, \citet{saa+10} and \citet{saa+12} hypothesize a stationary 
shell detached from the ejecta. The shell is accelerated to a constant speed $v_{sh}$ by colliding with fast-moving ejecta with velocities greater than $v_{sh}$. 
In fact, some simulations of WD mergers show that the outermost material forms such a stationary envelope that collides with fast-moving ejecta \citep{hk96}. 
Following the calculation procedure in \citet{saa+10} and \citet{saa+12}, we derive an envelope mass of $0.15^{+0.11}_{-0.01}\sm$ for iPTF13asv. 
This shell increases the total mass of the system to $1.59^{+0.45}_{-0.12}\sm$
The $^{56}Ni$, shell, and total masses of iPTF13asv are similar to those derived for SN20080522-000 in \citet{saa+12}. 

The detached shell has little effect on the $\gamma$-ray opacity and peak luminosity of an SN. Since the rise time is proportional to $M_{tot}^{1/2}$, the
massless detached shell will not make the SN rise substantially longer than usual. 

%Although the detached shell has little effect on the $\gamma$-ray opacity and peak luminosity of a SN, it reddens the SN and extends its rise time. 
%However, compared to other super-Chandrasekhar events \citep{saa+12}, iPTF13asv shows a blue intrinsic color and a much shorter rise time. 
%Based on the color and rise time, we infer that iPTF13asv is unlikely to have this detached shell. 
%

\section{Discussions}
\label{sec:discussion}

\subsection{Origin of Strong UV Emission}
Strong UV emission in an SN \RNum{1}a may be powered by an extrinsic SN-companion collision \citep{Kasen2010}. 
In fact, in the UV-luminous SN2011de, \citet{Brown2014} offered a possible explanation for its \textit{Swift} light curve as a collision between the SN ejecta
and a companion star. Here we consider the same model to interpret the strong UV emission of iPTF13asv. 

\begin{figure}
\centering
\includegraphics[width=0.45\textwidth]{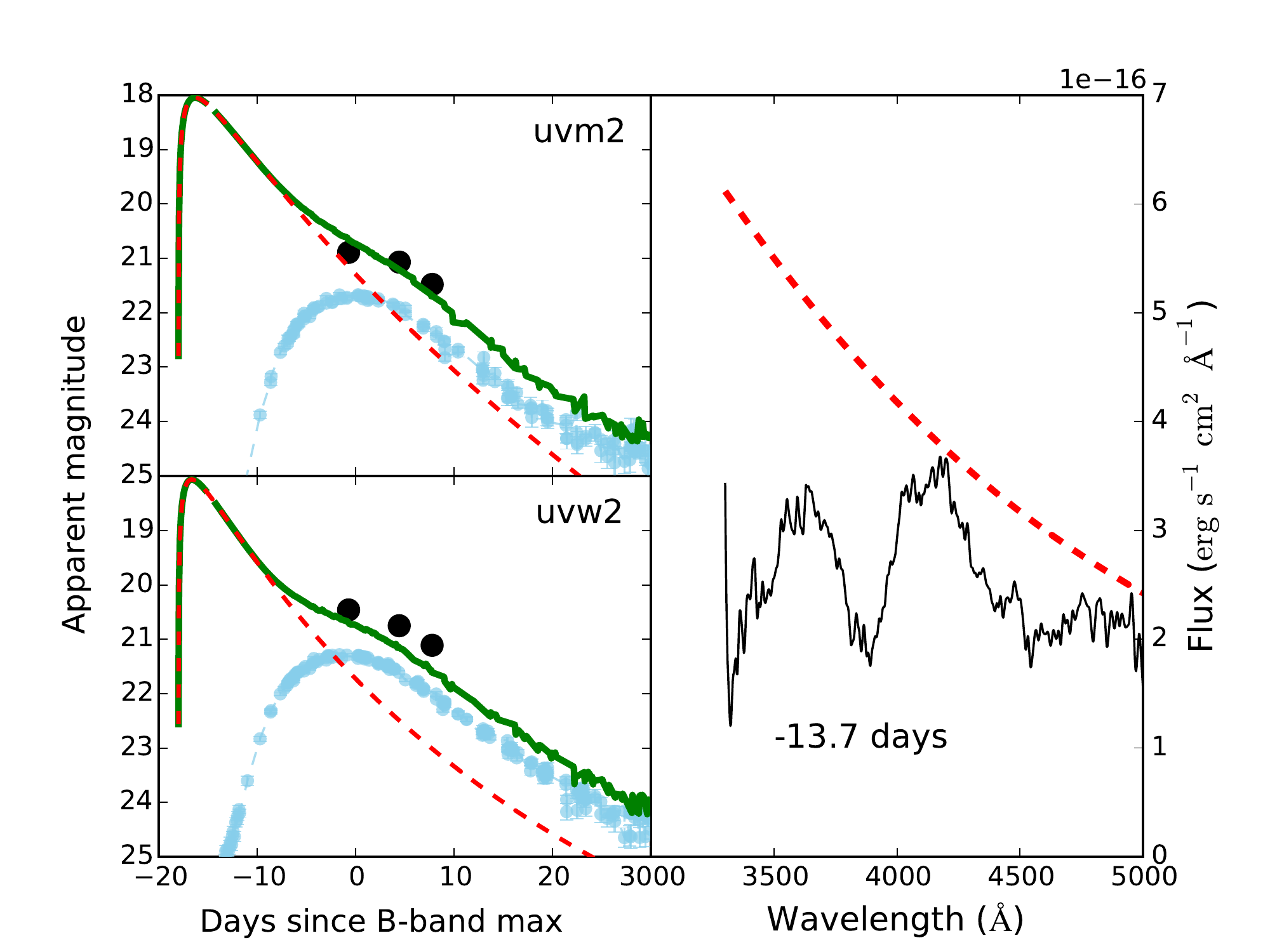}
\caption{Comparison between the SN-companion interaction signature model and iPTF13asv data. \textit{Left}: the $uvm2$ and
$uvw2$ light curves of iPTF13asv (black circles) are compared to the total light curves (green) which combine the SN-companion interaction component
(red; \citealt{Kasen2010}) and the SN intrinsic emission (cyan). \textit{Right}: the iPTF13asv spectrum at $-13.7$ days (black) is compared to 
the thermal spectrum of SN-companion collision (red). 
\label{fig:SN-companion}}
\end{figure}

We utilize the scaling relation in \citet{Kasen2010} to fit the observed $uvm2$ light curve. In order to account for the non-negligible emission 
from the SN itself, we use the well-sampled $uvm2$ light curve of SN2011fe as a template. The fitting result shows that the companion star
is located at $2\times10^{13}\,\rm{cm}$ away from the exploding WD (left panel of Figure \ref{fig:SN-companion}). Given a typical mass ratio of a few, 
the companion has a radius of $\sim100\sr$ and fills its Roche lobe. 

Although the model fit to the UV light curve looks plausible, it overpredicts the SN emission at very early phases. In the R-band light curve within a few days of explosion, 
the model-predicted SN flux ($200\rm{\mu Jy}$) is higher than the observed fluxes (the inset of Figure \ref{fig:lightCurve}) by a factor of $>30\%$. At $-13.7$ days, 
the predicted thermal emission flux from the model below 4000\,\AA\ is also much higher than the observed spectrum (right panel of Figure \ref{fig:SN-companion}). 
Hence, we conclude that the strong UV emission 
seen in iPTF13asv is not produced by SN-companion collision. 

As a result of the above analysis, we are forced to conclude that the strong UV emission is intrinsic. In fact, the strong UV emission and the lack of iron in 
early-phase spectra are probably causally related, as the iron group elements are the major absorbers of UV photons. These two observational facts, together
with the near-IR secondary peak, strongly suggest that iPTF13asv has a stratified ejecta along the line of sight, with strong concentration of iron group 
elements near the center of the explosion. 
%
%The latter implies an ejecta structure of weak mixing with most iron group elements concentrated in the low-velocity
%zones of the ejecta. Since the line blanketing of iron group elements dominates the UV opacity, the optical opacity of UV photons is very small until the
%photosphere recedes into the low-velocity zones of the ejecta. Consequently, the SN appears to be more UV luminous around maximum. 
%
%The concentration of iron group elements is also supported by the strong secondary peak in the near-IR. The near-IR secondary peak results from transition 
%of iron group elements from doubly ionized states to singly ionized states. The double-peaked morphology of the near-IR light curve is therefore a direct consequence 
%of the abundance stratification of the ejecta, in particular, the concentration of iron group elements in the central regions \citep{Kasen2006}. 
%
%To sum up, the strong UV emission, the absence of iron in early spectra and the near-IR secondary maximum together
%suggest that iPTF13asv has a stratified ejecta with strong concentration of iron group elements in the center. 

\subsection{iPTF13asv as an Intermediate Case between Normal and Super-Chandrasekhar Subclasses}
In Table \ref{tab:comparison}, we summarize a comparison of normal SNe, iPTF13asv, and super-Chandrasekhar SNe. 
As can be seen from the table, on the one hand, iPTF13asv shares similar light curve shapes and near-IR secondary peak with normal events. SNID also finds decent 
spectral matches between iPTF13asv and normal events. On the other hand, the peak radiation of iPTF13asv is as bright as super-Chandrasekhar events in both optical and UV. 
The evolution of \ion{Si}{2} velocities of iPTF13asv is also similar to those of super-Chandrasekhar events. In addition, we derived an total ejecta mass slightly beyond
the Chandrasekhar mass limit. Hence, we classify iPTF13asv as an intermediate case between normal and super-Chandrasekhar subclasses. 

In addition to the features listed in the table, the H-band break, a sharp spectral feature formed by absorption of \ion{Fe}{2}, \ion{Co}{2} and \ion{Ni}{2} \citep{hmp+13}, 
is also distinctive between super-Chandrasekhar and normal events. The H-band break emerges around the maximum for normal SNe and decays to disappear within 
a month of maximum. In contrast, this feature does not appear in the super-Chandrasekhar events. However, the only near-IR spectrum of iPTF13asv is taken one 
month after the maximum. Therefore, we cannot determine whether iPTF13asv shows the H-band break or not.

\begin{deluxetable*}{ccccc}
\tablecolumns{5}
\tablewidth{0pt}
\tablecaption{Comparison of iPTF13asv to Normal and Super-Chandrasekhar SNe\label{tab:comparison}}
\tablehead{
	\colhead{Feature}   &   \colhead{Normal}   &  \colhead{iPTF13asv}    &    \colhead{Super-Chandrasekhar}    &   \colhead{Section\tablenotemark{a}}
}
\startdata
B-band absolute peak magnitude   &  $-18$ -- $-19.5$  &  $-19.97\pm0.06$   &  $<-19.6$  &  \S\ref{sec:analysis:lightcurve}\\
UV absolute peak magnitude    &  $\gtrsim-15$  &  $-15.25$  &   $\lesssim-15$  & \S\ref{sec:analysis:lightcurve} and Figure \ref{fig:swiftIaSample}\\
B-band $\Delta m_{15}$ (mag)    & $0.8$ -- $1.2$   &  $1.0$   &  $\simeq0.6$   & \S\ref{sec:analysis:lightcurve} \\
Near-IR secondary peak   &   strong  &   strong   &   weak or absent  & \S\ref{sec:analysis:lightcurve} and Figure \ref{fig:lcComparison} \\
SNID                                  &  normal   &  normal   & super-Chandrasekhar  &  \S\ref{sec:snid} and Table \ref{tab:snid}\\
Carbon feature after max   &  no   &  weak   &  strong  & \S\ref{sec:carbon} and Figures \ref{fig:carbon} and \ref{fig:carbonComparison} \\
\ion{Si}{2}\,6355 velocity at max ($10^3\,\rm{km}\,\rm{s}^{-1}$)\tablenotemark{b}  & $10$ -- $14$   &   $10$    &   $8$ -- $12$    & \S\ref{sec:velocity} and Figure \ref{fig:velocity} \\
\ion{Si}{2}\,6355 velocity gradient at max  ($\rm{km}\,\rm{s}^{-1}\,\rm{day}^{-1}$)\tablenotemark{b}   &  $50$ -- $150$   &  $\sim0$   &   $-72$ -- $46$  & \S\ref{sec:velocity} and Figure \ref{fig:velocityGradient} \\
$^{56}$Ni mass ($\sm$)\tablenotemark{c}    &  $0.3$ -- $0.6$   & $0.81^{+0.10}_{-0.18}$   &  $>0.75$   &  \S\ref{sec:mass:mass} \\
Ejecta mass ($\sm$)\tablenotemark{c}    &  $0.8$ -- $1.5$     &   $1.59^{+0.45}_{-0.12}$   &   $>1.5$  & \S\ref{sec:mass:mass}
\enddata
\tablenotetext{a}{This column points to sections in this paper that discuss corresponding features}
\tablenotetext{b}{The velocity measurements of normal events are from \citet{fsk11}. Those of super-Chandrasekhar events are from \citet{saa+12}. }
\tablenotetext{c}{The mass measurements of normal events are from \citet{sra14}. Those of super-Chandrasekhar events are from \citet{saa+12}.}
\end{deluxetable*}

\subsection{Progenitor}
\label{sec:discussion:progenitor}

The massive ejecta and the stratification of the ejecta favor a DD progenitor system for iPTF13asv.
In an SD system, a non-rotating WD cannot exceed the Chandrasekhar mass limit, and it is not clear in reality how rotation could
increase this mass limit. Hydrodynamic simulations (e.g., \citealt{KSF2010,ssk+13}) also show that explosions in an SD system cannot avoid 
a certain level of mixing in the ejecta. Hence, these models do not easily concentrate most of the iron group elements in the center of the ejecta. 
For merging WDs, in contrast, simulations of prompt detonation (e.g., \citealt{mrk+14}) produce strongly stratified structures along polar directions in
asymmetric ejecta. In these directions, iron group elements are confined to the low-velocity regions. 

We also consider the core-degenerate scenario to explain the progenitor system of iPTF13asv (e.g., \citealt{sga14}), but more exploration in this scenario
is needed to explain the weak mixing of iron group elements in the fast-moving ejecta, persistent carbon features after maximum, 
and low but almost constant Si II velocity. 

\subsection{iPTF13asv in cosmology}
\label{sec:cosmology}

\begin{figure}
\includegraphics[width=0.45\textwidth]{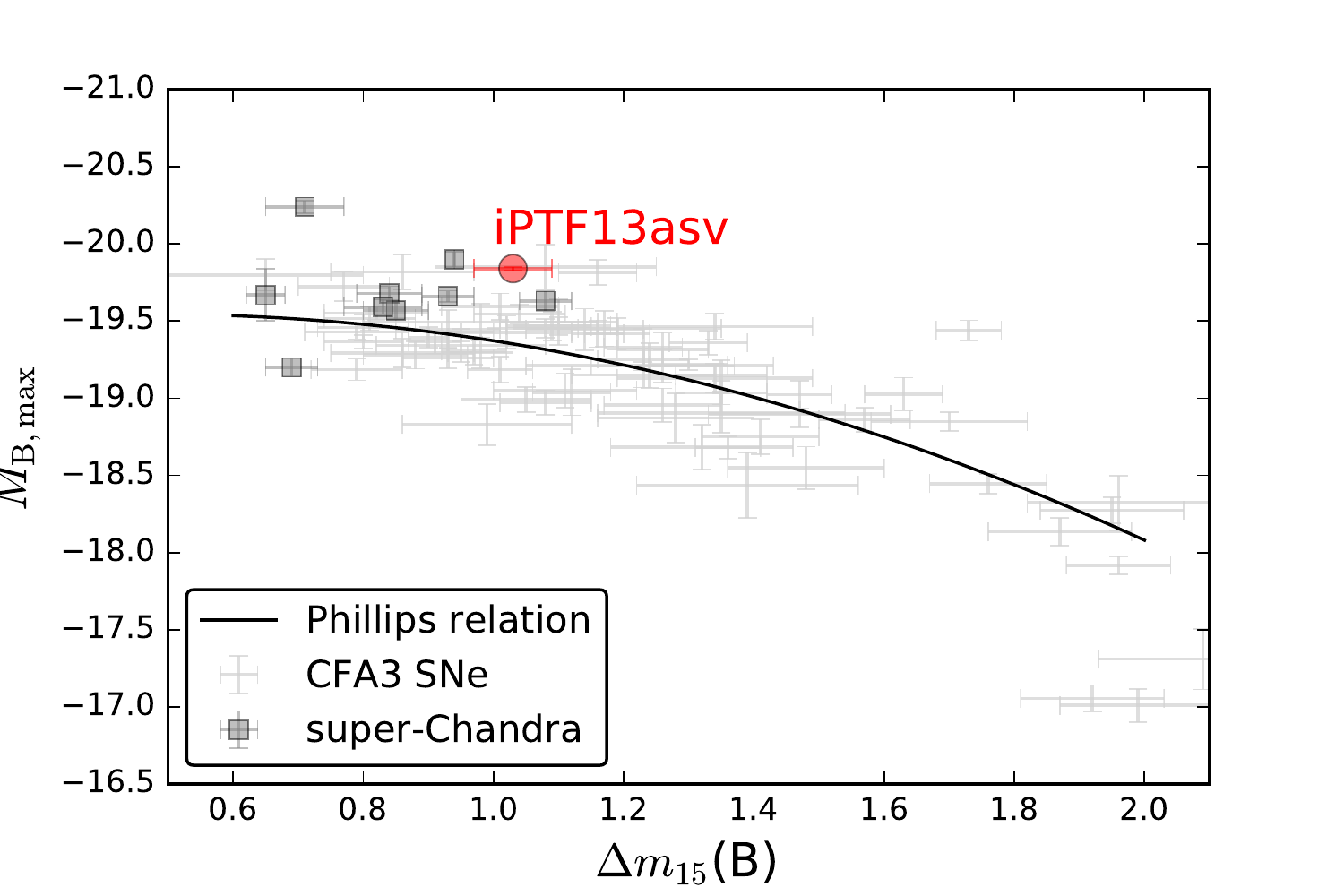}
\caption{iPTF13asv is an outlier from the the Phillips relation. The Phillips relation is taken from \citet{pls+99}. The CfA3 data are from
\citet{hck+12}. The super-Chandrasekhar events are from \citet{saa+10} and \citet{saa+12}. 
\label{fig:phillips}}
\end{figure}

Unsurprisingly, iPTF13asv is an outlier from the Phillips relation \citep{Phillips1993}. As shown in Figure \ref{fig:phillips}, iPTF13asv is above the
empirical relation by half a magnitude. 

To have better calibration in cosmology, a third color term is introduced in the Phillips relation, i.e., 
\begin{equation}
\mu=m_B^*-(M_B-\alpha x_1+\beta c)\ ,
\end{equation}
where $\mu$ is the distance modulus; $m_B^*$ is the observed peak magnitude in the rest-frame B band; $\alpha$, $\beta$ and $M_B$ are free
parameters. To account for the dependence on the host galaxy properties, \citet{SGC2011} suggest to use different values of $M_B$ for galaxies of stellar mass
greater than and less than $10^{10}\sm$. In the case of iPTF13asv, the stellar mass of its host galaxy is $\sim10^{7.8}\sm$. 
For galaxies with stellar mass less than $10^{10}\sm$, \citet{bkg+14} used a fiducial value of $H_0=70\,\rm{km}\,\rm{s}^{-1}\,\rm{Mpc}^{-1}$ and obtained $M_B=-19.04\pm0.01$, 
$\alpha=0.141\pm0.006$ and $\beta=3.101\pm0.075$. Using the same $H_0$ and the iPTF13asv measurements of $m_B^*=16.28\pm0.03$, $x_1=0.37\pm0.09$ and 
$c=-0.16\pm0.02$, we find that iPTF13asv can still be included in this empirical relation and thus be useful for cosmographic measurements, whereas super-Chandrasekhar 
events are outliers of this empirical relation \citep{saa+12}.

\subsection{UV-luminous SNe at High Redshifts}

Spectroscopic classification for high-redshift SNe requires very long integration on big telescopes. 
Therefore, in high-redshift SN surveys, an optical-UV ``dropout'' is introduced to preselect type \RNum{1}a
candidates. For example, \citet{rst+04, rsc+07} used the F850LP, F775W, and F606 filters on the {\it HST} 
Advanced Camera for Surveys (ACS) to search for SNe at redshifts up to $1.8$. 

\begin{figure}
\includegraphics[width=0.45\textwidth]{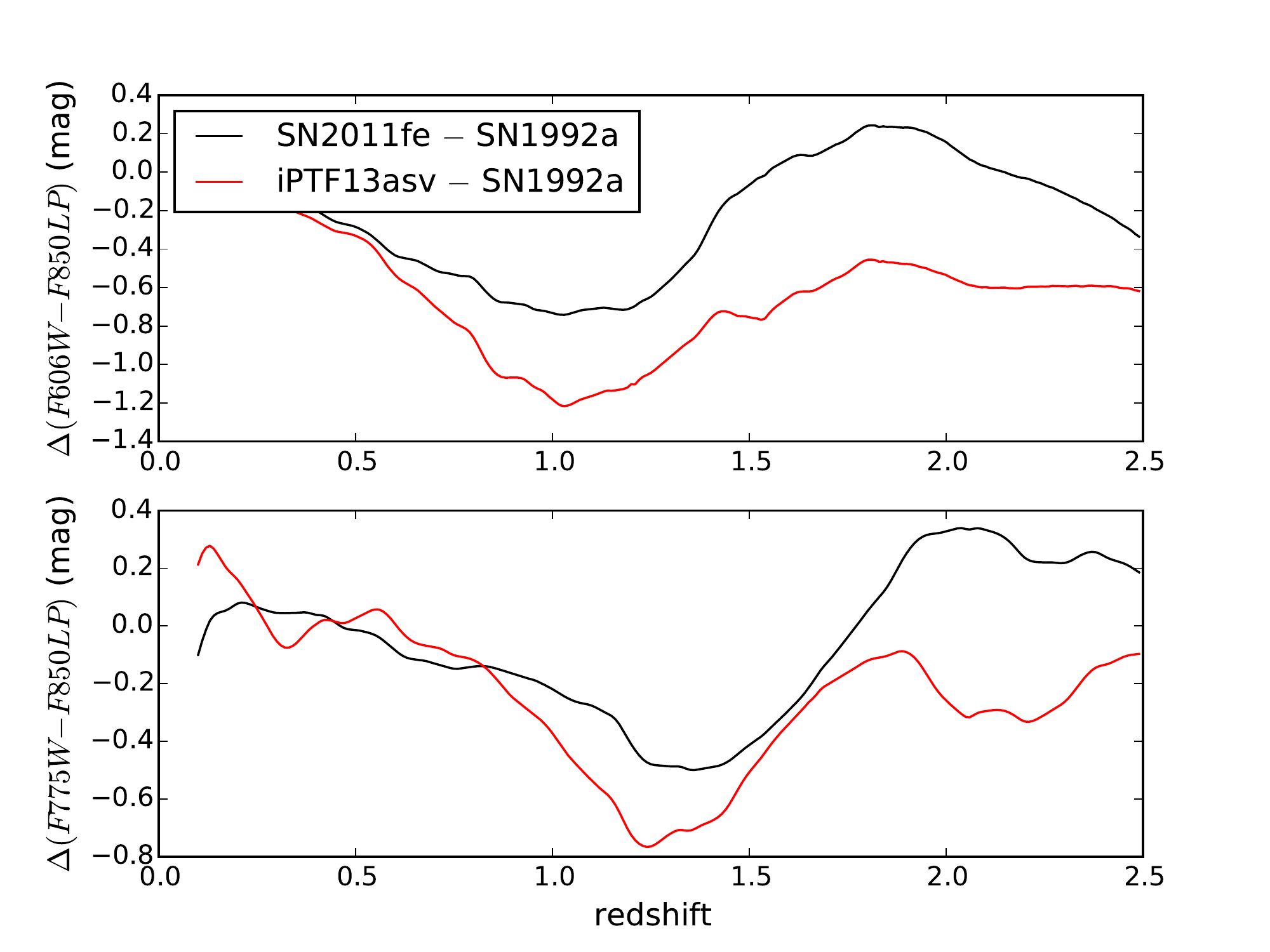}
\caption{Color difference as a function of redshift for three different SNe Ia.
\label{fig:color}}
\end{figure}

The color preselection criteria may introduce bias by ignoring UV-luminous SNe. In Figure \ref{fig:color}, we calculate the 
color difference in the F850LP, F775W, and F606 filters for normal SN1992a \citep{kjl+93}, near-UV blue SN2011fe \citep{MSH2014}), 
and UV-luminous iPTF13asv at different redshifts. As can be seen, the three SNe show different colors at high redshifts. Although
the rate of iPTF13asv-like events is probably low in the nearby Universe, there might be more such SNe at high redshifts as
there are more metal-poor dwarf galaxies at high redshifts. Hence it might become a non-negligible component in estimating the
SN rate at high redshifts. 

%Since the UV-luminous SNe Ia like iPTF13asv are useful for 
%cosmological distance measurements, including these SNe in future surveys will increase the sample size for cosmography. 

%
% In order to reconstruct the UV spectrum of iPTF13asv at maximum, we use the spectrum of 
%SN2011fe below 2800\,\AA\ (where the UV ``dropout'' is), rescale it to match the $uvm2$ magnitude of iPTF13asv, and connect to the observed optical spectrum of iPTF13asv. 
%As can be seen in Figure \ref{fig:color}, the colors of F606W$-$F850LP and F775W$-$F850LP varies by at least one magnitude in the
%redshift range between $1.0$ and $1.8$. Such a large color variation suggest that the color selection criterion for high-redshift SNe Ia may have missed 
%UV-luminous events. Though UV-luminous SNe Ia are not very common in the nearby Universe, if they tend to appear more frequently
%in dwarf galaxies, then the fraction of missed UV-luminous SNe Ia might will increase at high redshifts. As a result, the current value of Type \RNum{1}a rate at 
%high redshift might be underestimated. 

\section{Conclusions}
\label{sec:conclusion}
In this paper, we present multi-wavelength observations of a peculiar overluminous Type \RNum{1}a supernova, iPTF13asv, discovered by
the intermediate Palomar Transient Factory. Although its light curve shape ($\Delta m_{15}=1.03\pm0.01$\,mag) and sharp secondary near-IR peak resemble characteristic features
of normal Type \RNum{1}a supernovae, iPTF13asv shows low but almost constant expansion velocities and persistent carbon absorption 
features after the maximum, both of which are commonly seen in super-Chandrasekhar events. We derive a $^{56}$Ni mass of $0.81^{+0.10}_{-0.18}\sm$ and a total
ejecta mass of $1.59^{+0.45}_{-0.12}\sm$. Therefore, we suggest that iPTF13asv is an intermediate case between the
normal and super-Chandrasekhar events. 

Our observations of iPTF13asv also show an absence of iron absorption features in its early-phase spectra until several days before maximum and strong UV
emission around peak. These observations, together with sharp near-IR secondary maxima, indicate that iPTF13asv has a stratified
structure along the line of sight, with synthesized iron group elements concentrated in the center of its ejecta. Compared to hydrodynamic simulations, 
only WD mergers might produce the inferred ejecta structure. Therefore, based on the stratified
ejecta and its similarity to super-Chandrasekhar events, we conclude that iPTF13asv originates from a double-degenerate progenitor system. 

We speculate that iPTF13asv might represent a transition case between normal and super-Chandrasekhar events. 
The current and upcoming time-domain surveys, such as DECam surveys, Zwicky Transient Facility, and LSST, will find many more Type \RNum{1}a supernovae
of different subclasses. Equipped with fast-turnaround follow-up observations which allow us to estimate the ejecta mass and the $^{56}$Ni mass, 
these surveys will map how spectral features (such as iron absorptions and carbon signatures) vary as a function of the total ejecta mass for supernovae.  
In particular, UV photometry and optical spectroscopy of these supernovae at early phases will reveal information about mixing of iron group elements. 
Understanding peculiar but possibly linking events like iPTF13asv in the context of large samples will further our knowledge about different subtypes of Type \RNum{1}a
supernovae and their physical origins.

\acknowledgments

We thank the anonymous referee for very useful comments and suggestions that substantially improved the manuscript. 
We are also grateful to M. Kromer for useful discussion about theoretic progenitor scenarios, and to A. De Cia, O. Yaron, 
D. Tal, D. Perley, K. Tinyanont, A. Waszczak, I. Arcavi, and S. Tang for performing observations and data reduction. 

This research is partly supported by the \textit{Swift} Guest Investigator program and by the National Science Foundation. 
Y.C. and M.M.K. acknowledge support from the National Science Foundation PIRE program grant 1545949. 
A.G. and R.A. acknowledge support from the Swedish Research Council and the Swedish Space Board.
E.Y.H. acknowledges the support provided by the Danish Agency for Science and Technology and Innovation
through a Sapere Aude Level 2 grant.

Some observations obtained with the SuperNova Integral Field Spectrograph on the University of Hawaii 2.2-m telescope as part 
of the Nearby Supernova Factory II project, a scientific collaboration between the Centre de Recherche Astronomique de Lyon, 
Institut de Physique Nucl'eaire de Lyon, Laboratoire de Physique Nucl'eaire et des Hautes Energies, Lawrence Berkeley National 
Laboratory, Yale University , University, University of Bonn , Max Planck Institute for Astrophysics, Tsinghua Center for Astrophysics, 
and Centre de Physique des Particules de Marseille.

Some data were obtained with ALFOSC, which is provided by
the Instituto de Astrofisica de Andalucia (IAA) under a joint agreement with the University
of Copenhagen and NOTSA. 

We also thank the RATIR project team and the staff of the Observatorio Astron\'{o}mico Nacional on
Sierra San Pedro M\'{a}rtir. RATIR is a collaboration between
the University of California, the Universidad Nacional
Aut\'{o}noma de M\'{e}xico, NASA Goddard Space Flight
Center, and Arizona State University, benefiting from the
loan of an H2RG detector and hardware and software support
from Teledyne Scientific and Imaging. RATIR, the
automation of the Harold L. Johnson Telescope of the
Observatorio Astron\'{o}mico Nacional on Sierra San Pedro
M\'{a}rtir, and the operation of both are funded through NASA
grants NNX09AH71G, NNX09AT02G, NNX10AI27G, and
NNX12AE66G, CONACyT grants LN 260369, and UNAM PAPIIT grant IG100414.

A portion of this work was carried out at the Jet Propulsion Laboratory, California Institute of Technology, 
under a contract with the National Aeronautics and Space Administration.  Copyright 2016 California Institute of Technology.  
All Rights Reserved. US Government Support Acknowledged.

\bibliography{ref}

\begin{thebibliography}{}
\expandafter\ifx\csname natexlab\endcsname\relax\def\natexlab#1{#1}\fi

\bibitem[{{Aldering} {et~al.}(2006){Aldering}, {Antilogus}, {Bailey}, {Baltay},
  {Bauer}, {Blanc}, {Bongard}, {Copin}, {Gangler}, {Gilles}, {Kessler},
  {Kocevski}, {Lee}, {Loken}, {Nugent}, {Pain}, {P{\'e}contal}, {Pereira},
  {Perlmutter}, {Rabinowitz}, {Rigaudier}, {Scalzo}, {Smadja}, {Thomas},
  {Wang}, {Weaver}, \& {Nearby Supernova Factory}}]{AAB2006}
{Aldering}, G., {Antilogus}, P., {Bailey}, S., {et~al.} 2006, \apj, 650, 510

\bibitem[{{Barone-Nugent} {et~al.}(2012){Barone-Nugent}, {Lidman}, {Wyithe},
  {Mould}, {Howell}, {Hook}, {Sullivan}, {Nugent}, {Arcavi}, {Cenko}, {Cooke},
  {Gal-Yam}, {Hsiao}, {Kasliwal}, {Maguire}, {Ofek}, {Poznanski}, \&
  {Xu}}]{blw+12}
{Barone-Nugent}, R.~L., {Lidman}, C., {Wyithe}, J.~S.~B., {et~al.} 2012,
  \mnras, 425, 1007

\bibitem[{{Betoule} {et~al.}(2014){Betoule}, {Kessler}, {Guy}, {Mosher},
  {Hardin}, {Biswas}, {Astier}, {El-Hage}, {Konig}, {Kuhlmann}, {Marriner},
  {Pain}, {Regnault}, {Balland}, {Bassett}, {Brown}, {Campbell}, {Carlberg},
  {Cellier-Holzem}, {Cinabro}, {Conley}, {D'Andrea}, {DePoy}, {Doi}, {Ellis},
  {Fabbro}, {Filippenko}, {Foley}, {Frieman}, {Fouchez}, {Galbany}, {Goobar},
  {Gupta}, {Hill}, {Hlozek}, {Hogan}, {Hook}, {Howell}, {Jha}, {Le Guillou},
  {Leloudas}, {Lidman}, {Marshall}, {M{\"o}ller}, {Mour{\~a}o}, {Neveu},
  {Nichol}, {Olmstead}, {Palanque-Delabrouille}, {Perlmutter}, {Prieto},
  {Pritchet}, {Richmond}, {Riess}, {Ruhlmann-Kleider}, {Sako}, {Schahmaneche},
  {Schneider}, {Smith}, {Sollerman}, {Sullivan}, {Walton}, \&
  {Wheeler}}]{bkg+14}
{Betoule}, M., {Kessler}, R., {Guy}, J., {et~al.} 2014, \aap, 568, A22

\bibitem[{{Blondin} \& {Tonry}(2007)}]{bt07}
{Blondin}, S., \& {Tonry}, J.~L. 2007, \apj, 666, 1024

\bibitem[{{Blondin} {et~al.}(2012){Blondin}, {Matheson}, {Kirshner}, {Mandel},
  {Berlind}, {Calkins}, {Challis}, {Garnavich}, {Jha}, {Modjaz}, {Riess}, \&
  {Schmidt}}]{bmk+12}
{Blondin}, S., {Matheson}, T., {Kirshner}, R.~P., {et~al.} 2012, \aj, 143, 126

\bibitem[{{Breeveld} {et~al.}(2011){Breeveld}, {Landsman}, {Holland}, {Roming},
  {Kuin}, \& {Page}}]{BLH2011}
{Breeveld}, A.~A., {Landsman}, W., {Holland}, S.~T., {et~al.} 2011, in American
  Institute of Physics Conference Series, Vol. 1358, American Institute of
  Physics Conference Series, ed. J.~E. {McEnery}, J.~L. {Racusin}, \&
  N.~{Gehrels}, 373--376

\bibitem[{{Brown}(2014)}]{Brown2014}
{Brown}, P.~J. 2014, \apjl, 796, L18

\bibitem[{{Brown} {et~al.}(2012){Brown}, {Dawson}, {de Pasquale}, {Gronwall},
  {Holland}, {Immler}, {Kuin}, {Mazzali}, {Milne}, {Oates}, \&
  {Siegel}}]{BDd2012}
{Brown}, P.~J., {Dawson}, K.~S., {de Pasquale}, M., {et~al.} 2012, \apj, 753,
  22

\bibitem[{{Brown} {et~al.}(2014){Brown}, {Kuin}, {Scalzo}, {Smitka}, {de
  Pasquale}, {Holland}, {Krisciunas}, {Milne}, \& {Wang}}]{bks+14}
{Brown}, P.~J., {Kuin}, P., {Scalzo}, R., {et~al.} 2014, \apj, 787, 29

\bibitem[{{Burns} {et~al.}(2014){Burns}, {Stritzinger}, {Phillips}, {Hsiao},
  {Contreras}, {Persson}, {Folatelli}, {Boldt}, {Campillay}, {Castell{\'o}n},
  {Freedman}, {Madore}, {Morrell}, {Salgado}, \& {Suntzeff}}]{bsp+14}
{Burns}, C.~R., {Stritzinger}, M., {Phillips}, M.~M., {et~al.} 2014, \apj, 789,
  32

\bibitem[{{Cao} {et~al.}(2015){Cao}, {Kulkarni}, {Howell}, {Gal-Yam},
  {Kasliwal}, {Valenti}, {Johansson}, {Amanullah}, {Goobar}, {Sollerman},
  {Taddia}, {Horesh}, {Sagiv}, {Cenko}, {Nugent}, {Arcavi}, {Surace},
  {Wo{\'z}niak}, {Moody}, {Rebbapragada}, {Bue}, \& {Gehrels}}]{ckh+15}
{Cao}, Y., {Kulkarni}, S.~R., {Howell}, D.~A., {et~al.} 2015, \nat, 521, 328

\bibitem[{{Cenko} {et~al.}(2006){Cenko}, {Fox}, {Moon}, {Harrison}, {Kulkarni},
  {Henning}, {Guzman}, {Bonati}, {Smith}, {Thicksten}, {Doyle}, {Petrie},
  {Gal-Yam}, {Soderberg}, {Anagnostou}, \& {Laity}}]{CFM2006}
{Cenko}, S.~B., {Fox}, D.~B., {Moon}, D.-S., {et~al.} 2006, \pasp, 118, 1396

\bibitem[{{Denicol{\'o}} {et~al.}(2002){Denicol{\'o}}, {Terlevich}, \&
  {Terlevich}}]{dtt02}
{Denicol{\'o}}, G., {Terlevich}, R., \& {Terlevich}, E. 2002, \mnras, 330, 69

\bibitem[{{Di Stefano} {et~al.}(2011){Di Stefano}, {Voss}, \& {Claeys}}]{11dvc}
{Di Stefano}, R., {Voss}, R., \& {Claeys}, J.~S.~W. 2011, \apjl, 738, L1

\bibitem[{{Filippenko} {et~al.}(1992){Filippenko}, {Richmond}, {Matheson},
  {Shields}, {Burbidge}, {Cohen}, {Dickinson}, {Malkan}, {Nelson}, {Pietz},
  {Schlegel}, {Schmeer}, {Spinrad}, {Steidel}, {Tran}, \& {Wren}}]{frm+92}
{Filippenko}, A.~V., {Richmond}, M.~W., {Matheson}, T., {et~al.} 1992, \apjl,
  384, L15

\bibitem[{{Firth} {et~al.}(2015){Firth}, {Sullivan}, {Gal-Yam}, {Howell},
  {Maguire}, {Nugent}, {Piro}, {Baltay}, {Feindt}, {Hadjiyksta}, {McKinnon},
  {Ofek}, {Rabinowitz}, \& {Walker}}]{FSG2015}
{Firth}, R.~E., {Sullivan}, M., {Gal-Yam}, A., {et~al.} 2015, \mnras, 446, 3895

\bibitem[{{Fitzpatrick}(1999)}]{ftz99}
{Fitzpatrick}, E.~L. 1999, \pasp, 111, 63

\bibitem[{{Foley} {et~al.}(2014){Foley}, {McCully}, {Jha}, {Bildsten}, {Fong},
  {Narayan}, {Rest}, \& {Stritzinger}}]{fmj+14}
{Foley}, R.~J., {McCully}, C., {Jha}, S.~W., {et~al.} 2014, \apj, 792, 29

\bibitem[{{Foley} {et~al.}(2011){Foley}, {Sanders}, \& {Kirshner}}]{fsk11}
{Foley}, R.~J., {Sanders}, N.~E., \& {Kirshner}, R.~P. 2011, \apj, 742, 89

\bibitem[{{Foster} {et~al.}(2012){Foster}, {Hopkins}, {Gunawardhana},
  {Lara-L{\'o}pez}, {Sharp}, {Steele}, {Taylor}, {Driver}, {Baldry}, {Bamford},
  {Liske}, {Loveday}, {Norberg}, {Peacock}, {Alpaslan}, {Bauer},
  {Bland-Hawthorn}, {Brough}, {Cameron}, {Colless}, {Conselice}, {Croom},
  {Frenk}, {Hill}, {Jones}, {Kelvin}, {Kuijken}, {Nichol}, {Owers},
  {Parkinson}, {Pimbblet}, {Popescu}, {Prescott}, {Robotham}, {Lopez-Sanchez},
  {Sutherland}, {Thomas}, {Tuffs}, {van Kampen}, \& {Wijesinghe}}]{fhg+12}
{Foster}, C., {Hopkins}, A.~M., {Gunawardhana}, M., {et~al.} 2012, \aap, 547,
  A79

\bibitem[{{Garavini} {et~al.}(2004){Garavini}, {Folatelli}, {Goobar}, {Nobili},
  {Aldering}, {Amadon}, {Amanullah}, {Astier}, {Balland}, {Blanc}, {Burns},
  {Conley}, {Dahl{\'e}n}, {Deustua}, {Ellis}, {Fabbro}, {Fan}, {Frye}, {Gates},
  {Gibbons}, {Goldhaber}, {Goldman}, {Groom}, {Haissinski}, {Hardin}, {Hook},
  {Howell}, {Kasen}, {Kent}, {Kim}, {Knop}, {Lee}, {Lidman}, {Mendez},
  {Miller}, {Moniez}, {Mour{\~a}o}, {Newberg}, {Nugent}, {Pain}, {Perdereau},
  {Perlmutter}, {Prasad}, {Quimby}, {Raux}, {Regnault}, {Rich}, {Richards},
  {Ruiz-Lapuente}, {Sainton}, {Schaefer}, {Schahmaneche}, {Smith}, {Spadafora},
  {Stanishev}, {Walton}, {Wang}, {Wood-Vasey}, \& {Supernova Cosmology
  Project}}]{GFG2004}
{Garavini}, G., {Folatelli}, G., {Goobar}, A., {et~al.} 2004, \aj, 128, 387

\bibitem[{{Goobar} \& {Leibundgut}(2011)}]{GL2011}
{Goobar}, A., \& {Leibundgut}, B. 2011, Annual Review of Nuclear and Particle
  Science, 61, 251

\bibitem[{{Goobar} {et~al.}(2015){Goobar}, {Kromer}, {Siverd}, {Stassun},
  {Pepper}, {Amanullah}, {Kasliwal}, {Sollerman}, \& {Taddia}}]{gks+15}
{Goobar}, A., {Kromer}, M., {Siverd}, R., {et~al.} 2015, \apj, 799, 106

\bibitem[{{Guy} {et~al.}(2007){Guy}, {Astier}, {Baumont}, {Hardin}, {Pain},
  {Regnault}, {Basa}, {Carlberg}, {Conley}, {Fabbro}, {Fouchez}, {Hook},
  {Howell}, {Perrett}, {Pritchet}, {Rich}, {Sullivan}, {Antilogus}, {Aubourg},
  {Bazin}, {Bronder}, {Filiol}, {Palanque-Delabrouille}, {Ripoche}, \&
  {Ruhlmann-Kleider}}]{GAB2007}
{Guy}, J., {Astier}, P., {Baumont}, S., {et~al.} 2007, \aap, 466, 11

\bibitem[{{Hachisu} {et~al.}(2012){Hachisu}, {Kato}, \& {Nomoto}}]{hkn+12}
{Hachisu}, I., {Kato}, M., \& {Nomoto}, K. 2012, \apjl, 756, L4

\bibitem[{{Hicken} {et~al.}(2007){Hicken}, {Garnavich}, {Prieto}, {Blondin},
  {DePoy}, {Kirshner}, \& {Parrent}}]{hgp+07}
{Hicken}, M., {Garnavich}, P.~M., {Prieto}, J.~L., {et~al.} 2007, \apjl, 669,
  L17

\bibitem[{{Hicken} {et~al.}(2012){Hicken}, {Challis}, {Kirshner}, {Rest},
  {Cramer}, {Wood-Vasey}, {Bakos}, {Berlind}, {Brown}, {Caldwell}, {Calkins},
  {Currie}, {de Kleer}, {Esquerdo}, {Everett}, {Falco}, {Fernandez},
  {Friedman}, {Groner}, {Hartman}, {Holman}, {Hutchins}, {Keys}, {Kipping},
  {Latham}, {Marion}, {Narayan}, {Pahre}, {Pal}, {Peters}, {Perumpilly},
  {Ripman}, {Sipocz}, {Szentgyorgyi}, {Tang}, {Torres}, {Vaz}, {Wolk}, \&
  {Zezas}}]{hck+12}
{Hicken}, M., {Challis}, P., {Kirshner}, R.~P., {et~al.} 2012, \apjs, 200, 12

\bibitem[{{Hoeflich} \& {Khokhlov}(1996)}]{hk96}
{Hoeflich}, P., \& {Khokhlov}, A. 1996, \apj, 457, 500

\bibitem[{{Howell} {et~al.}(2005){Howell}, {Sullivan}, {Perrett}, {Bronder},
  {Hook}, {Astier}, {Aubourg}, {Balam}, {Basa}, {Carlberg}, {Fabbro},
  {Fouchez}, {Guy}, {Lafoux}, {Neill}, {Pain}, {Palanque-Delabrouille},
  {Pritchet}, {Regnault}, {Rich}, {Taillet}, {Knop}, {McMahon}, {Perlmutter},
  \& {Walton}}]{HSP2005}
{Howell}, D.~A., {Sullivan}, M., {Perrett}, K., {et~al.} 2005, \apj, 634, 1190

\bibitem[{{Howell} {et~al.}(2006){Howell}, {Sullivan}, {Nugent}, {Ellis},
  {Conley}, {Le Borgne}, {Carlberg}, {Guy}, {Balam}, {Basa}, {Fouchez}, {Hook},
  {Hsiao}, {Neill}, {Pain}, {Perrett}, \& {Pritchet}}]{hsn+06}
{Howell}, D.~A., {Sullivan}, M., {Nugent}, P.~E., {et~al.} 2006, \nat, 443, 308

\bibitem[{{Hsiao} {et~al.}(2007){Hsiao}, {Conley}, {Howell}, {Sullivan},
  {Pritchet}, {Carlberg}, {Nugent}, \& {Phillips}}]{hch+07}
{Hsiao}, E.~Y., {Conley}, A., {Howell}, D.~A., {et~al.} 2007, \apj, 663, 1187

\bibitem[{{Hsiao} {et~al.}(2013){Hsiao}, {Marion}, {Phillips}, {Burns},
  {Winge}, {Morrell}, {Contreras}, {Freedman}, {Kromer}, {Gall}, {Gerardy},
  {H{\"o}flich}, {Im}, {Jeon}, {Kirshner}, {Nugent}, {Persson}, {Pignata},
  {Roth}, {Stanishev}, {Stritzinger}, \& {Suntzeff}}]{hmp+13}
{Hsiao}, E.~Y., {Marion}, G.~H., {Phillips}, M.~M., {et~al.} 2013, \apj, 766,
  72

\bibitem[{{Jeffery}(1999)}]{j99}
{Jeffery}, D.~J. 1999, ArXiv Astrophysics e-prints, astro-ph/9907015

\bibitem[{{Jeffery} {et~al.}(2006){Jeffery}, {Branch}, \& {Baron}}]{jbb06}
{Jeffery}, D.~J., {Branch}, D., \& {Baron}, E. 2006, ArXiv Astrophysics
  e-prints, astro-ph/0609804

\bibitem[{{Justham}(2011)}]{j11}
{Justham}, S. 2011, \apjl, 730, L34

\bibitem[{{Kasen}(2006)}]{Kasen2006}
{Kasen}, D. 2006, \apj, 649, 939

\bibitem[{{Kasen}(2010)}]{Kasen2010}
---. 2010, \apj, 708, 1025

\bibitem[{{Kelly} {et~al.}(2014){Kelly}, {Fox}, {Filippenko}, {Cenko}, {Prato},
  {Schaefer}, {Shen}, {Zheng}, {Graham}, \& {Tucker}}]{kff+14}
{Kelly}, P.~L., {Fox}, O.~D., {Filippenko}, A.~V., {et~al.} 2014, \apj, 790, 3

\bibitem[{{Kennicutt}(1998)}]{Kennicutt1998}
{Kennicutt}, Jr., R.~C. 1998, \araa, 36, 189

\bibitem[{{Kirshner} {et~al.}(1993){Kirshner}, {Jeffery}, {Leibundgut},
  {Challis}, {Sonneborn}, {Phillips}, {Suntzeff}, {Smith}, {Winkler}, {Winge},
  {Hamuy}, {Hunter}, {Roth}, {Blades}, {Branch}, {Chevalier}, {Fransson},
  {Panagia}, {Wagoner}, {Wheeler}, \& {Harkness}}]{kjl+93}
{Kirshner}, R.~P., {Jeffery}, D.~J., {Leibundgut}, B., {et~al.} 1993, \apj,
  415, 589

\bibitem[{{Kriek} {et~al.}(2009){Kriek}, {van Dokkum}, {Labb{\'e}}, {Franx},
  {Illingworth}, {Marchesini}, \& {Quadri}}]{kvl+09}
{Kriek}, M., {van Dokkum}, P.~G., {Labb{\'e}}, I., {et~al.} 2009, \apj, 700,
  221

\bibitem[{{Krisciunas} {et~al.}(2000){Krisciunas}, {Hastings}, {Loomis},
  {McMillan}, {Rest}, {Riess}, \& {Stubbs}}]{khl+00}
{Krisciunas}, K., {Hastings}, N.~C., {Loomis}, K., {et~al.} 2000, \apj, 539,
  658

\bibitem[{{Kromer} {et~al.}(2010){Kromer}, {Sim}, {Fink}, {R{\"o}pke},
  {Seitenzahl}, \& {Hillebrandt}}]{KSF2010}
{Kromer}, M., {Sim}, S.~A., {Fink}, M., {et~al.} 2010, \apj, 719, 1067

\bibitem[{{Kushnir} {et~al.}(2013){Kushnir}, {Katz}, {Dong}, {Livne}, \&
  {Fern{\'a}ndez}}]{kkd+13}
{Kushnir}, D., {Katz}, B., {Dong}, S., {Livne}, E., \& {Fern{\'a}ndez}, R.
  2013, \apjl, 778, L37

\bibitem[{{Lantz} {et~al.}(2004){Lantz}, {Aldering}, {Antilogus}, {Bonnaud},
  {Capoani}, {Castera}, {Copin}, {Dubet}, {Gangler}, {Henault}, {Lemonnier},
  {Pain}, {Pecontal}, {Pecontal}, \& {Smadja}}]{SNIFS}
{Lantz}, B., {Aldering}, G., {Antilogus}, P., {et~al.} 2004, in Society of
  Photo-Optical Instrumentation Engineers (SPIE) Conference Series, Vol. 5249,
  Optical Design and Engineering, ed. L.~{Mazuray}, P.~J. {Rogers}, \&
  R.~{Wartmann}, 146--155

\bibitem[{{Law} {et~al.}(2009){Law}, {Kulkarni}, {Dekany}, {Ofek}, {Quimby},
  {Nugent}, {Surace}, {Grillmair}, {Bloom}, {Kasliwal}, {Bildsten}, {Brown},
  {Cenko}, {Ciardi}, {Croner}, {Djorgovski}, {van Eyken}, {Filippenko}, {Fox},
  {Gal-Yam}, {Hale}, {Hamam}, {Helou}, {Henning}, {Howell}, {Jacobsen},
  {Laher}, {Mattingly}, {McKenna}, {Pickles}, {Poznanski}, {Rahmer}, {Rau},
  {Rosing}, {Shara}, {Smith}, {Starr}, {Sullivan}, {Velur}, {Walters}, \&
  {Zolkower}}]{LKD2009}
{Law}, N.~M., {Kulkarni}, S.~R., {Dekany}, R.~G., {et~al.} 2009, \pasp, 121,
  1395

\bibitem[{{Li} {et~al.}(2011){Li}, {Bloom}, {Podsiadlowski}, {Miller}, {Cenko},
  {Jha}, {Sullivan}, {Howell}, {Nugent}, {Butler}, {Ofek}, {Kasliwal},
  {Richards}, {Stockton}, {Shih}, {Bildsten}, {Shara}, {Bibby}, {Filippenko},
  {Ganeshalingam}, {Silverman}, {Kulkarni}, {Law}, {Poznanski}, {Quimby},
  {McCully}, {Patel}, {Maguire}, \& {Shen}}]{lbp+11}
{Li}, W., {Bloom}, J.~S., {Podsiadlowski}, P., {et~al.} 2011, \nat, 480, 348

\bibitem[{{Lundqvist} {et~al.}(2015){Lundqvist}, {Nyholm}, {Taddia},
  {Sollerman}, {Johansson}, {Kozma}, {Lundqvist}, {Fransson}, {Garnavich},
  {Kromer}, {Shappee}, \& {Goobar}}]{lnt+15}
{Lundqvist}, P., {Nyholm}, A., {Taddia}, F., {et~al.} 2015, \aap, 577, A39

\bibitem[{{Maeda} \& {Iwamoto}(2009)}]{mi09}
{Maeda}, K., \& {Iwamoto}, K. 2009, \mnras, 394, 239

\bibitem[{{Maoz} {et~al.}(2014){Maoz}, {Mannucci}, \& {Nelemans}}]{mmn14}
{Maoz}, D., {Mannucci}, F., \& {Nelemans}, G. 2014, \araa, 52, 107

\bibitem[{{Margutti} {et~al.}(2014){Margutti}, {Parrent}, {Kamble},
  {Soderberg}, {Foley}, {Milisavljevic}, {Drout}, \& {Kirshner}}]{mpk+14}
{Margutti}, R., {Parrent}, J., {Kamble}, A., {et~al.} 2014, \apj, 790, 52

\bibitem[{{Marion} {et~al.}(2016){Marion}, {Brown}, {Vink{\'o}}, {Silverman},
  {Sand}, {Challis}, {Kirshner}, {Wheeler}, {Berlind}, {Brown}, {Calkins},
  {Camacho}, {Dhungana}, {Foley}, {Friedman}, {Graham}, {Howell}, {Hsiao},
  {Irwin}, {Jha}, {Kehoe}, {Macri}, {Maeda}, {Mandel}, {McCully}, {Pandya},
  {Rines}, {Wilhelmy}, \& {Zheng}}]{mbv+16}
{Marion}, G.~H., {Brown}, P.~J., {Vink{\'o}}, J., {et~al.} 2016, \apj, 820, 92

\bibitem[{{Matheson} {et~al.}(2008){Matheson}, {Kirshner}, {Challis}, {Jha},
  {Garnavich}, {Berlind}, {Calkins}, {Blondin}, {Balog}, {Bragg}, {Caldwell},
  {Dendy Concannon}, {Falco}, {Graves}, {Huchra}, {Kuraszkiewicz}, {Mader},
  {Mahdavi}, {Phelps}, {Rines}, {Song}, \& {Wilkes}}]{mkc+08}
{Matheson}, T., {Kirshner}, R.~P., {Challis}, P., {et~al.} 2008, \aj, 135, 1598

\bibitem[{{Mazzali} {et~al.}(1995){Mazzali}, {Danziger}, \&
  {Turatto}}]{MDT1995}
{Mazzali}, P.~A., {Danziger}, I.~J., \& {Turatto}, M. 1995, \aap, 297, 509

\bibitem[{{Mazzali} {et~al.}(2014){Mazzali}, {Sullivan}, {Hachinger}, {Ellis},
  {Nugent}, {Howell}, {Gal-Yam}, {Maguire}, {Cooke}, {Thomas}, {Nomoto}, \&
  {Walker}}]{MSH2014}
{Mazzali}, P.~A., {Sullivan}, M., {Hachinger}, S., {et~al.} 2014, \mnras, 439,
  1959

\bibitem[{{McCully} {et~al.}(2014){McCully}, {Jha}, {Foley}, {Bildsten},
  {Fong}, {Kirshner}, {Marion}, {Riess}, \& {Stritzinger}}]{mjf+14}
{McCully}, C., {Jha}, S.~W., {Foley}, R.~J., {et~al.} 2014, \nat, 512, 54

\bibitem[{{Milne} {et~al.}(2013){Milne}, {Brown}, {Roming}, {Bufano}, \&
  {Gehrels}}]{MBR2013}
{Milne}, P.~A., {Brown}, P.~J., {Roming}, P.~W.~A., {Bufano}, F., \& {Gehrels},
  N. 2013, \apj, 779, 23

\bibitem[{{Moll} {et~al.}(2014){Moll}, {Raskin}, {Kasen}, \&
  {Woosley}}]{mrk+14}
{Moll}, R., {Raskin}, C., {Kasen}, D., \& {Woosley}, S.~E. 2014, \apj, 785, 105

\bibitem[{{Nadyozhin}(1994)}]{n94}
{Nadyozhin}, D.~K. 1994, \apjs, 92, 527

\bibitem[{{Nobili} \& {Goobar}(2008)}]{ng08}
{Nobili}, S., \& {Goobar}, A. 2008, \aap, 487, 19

\bibitem[{{Nomoto} \& {Iben}(1985)}]{ni85}
{Nomoto}, K., \& {Iben}, Jr., I. 1985, \apj, 297, 531

\bibitem[{{Nugent} {et~al.}(2002){Nugent}, {Kim}, \& {Perlmutter}}]{nkp02}
{Nugent}, P., {Kim}, A., \& {Perlmutter}, S. 2002, \pasp, 114, 803

\bibitem[{{Nugent} {et~al.}(2011){Nugent}, {Sullivan}, {Cenko}, {Thomas},
  {Kasen}, {Howell}, {Bersier}, {Bloom}, {Kulkarni}, {Kandrashoff},
  {Filippenko}, {Silverman}, {Marcy}, {Howard}, {Isaacson}, {Maguire},
  {Suzuki}, {Tarlton}, {Pan}, {Bildsten}, {Fulton}, {Parrent}, {Sand},
  {Podsiadlowski}, {Bianco}, {Dilday}, {Graham}, {Lyman}, {James}, {Kasliwal},
  {Law}, {Quimby}, {Hook}, {Walker}, {Mazzali}, {Pian}, {Ofek}, {Gal-Yam}, \&
  {Poznanski}}]{NSC2011}
{Nugent}, P.~E., {Sullivan}, M., {Cenko}, S.~B., {et~al.} 2011, \nat, 480, 344

\bibitem[{{Oke} \& {Gunn}(1982)}]{DBSP}
{Oke}, J.~B., \& {Gunn}, J.~E. 1982, \pasp, 94, 586

\bibitem[{{Pan} {et~al.}(2014){Pan}, {Sullivan}, {Maguire}, {Hook}, {Nugent},
  {Howell}, {Arcavi}, {Botyanszki}, {Cenko}, {DeRose}, {Fakhouri}, {Gal-Yam},
  {Hsiao}, {Kulkarni}, {Laher}, {Lidman}, {Nordin}, {Walker}, \&
  {Xu}}]{PSM2014}
{Pan}, Y.-C., {Sullivan}, M., {Maguire}, K., {et~al.} 2014, \mnras, 438, 1391

\bibitem[{{Parrent} {et~al.}(2011){Parrent}, {Thomas}, {Fesen}, {Marion},
  {Challis}, {Garnavich}, {Milisavljevic}, {Vink{\`o}}, \& {Wheeler}}]{ptf+11}
{Parrent}, J.~T., {Thomas}, R.~C., {Fesen}, R.~A., {et~al.} 2011, \apj, 732, 30

\bibitem[{{Patat} {et~al.}(2007){Patat}, {Chandra}, {Chevalier}, {Justham},
  {Podsiadlowski}, {Wolf}, {Gal-Yam}, {Pasquini}, {Crawford}, {Mazzali},
  {Pauldrach}, {Nomoto}, {Benetti}, {Cappellaro}, {Elias-Rosa}, {Hillebrandt},
  {Leonard}, {Pastorello}, {Renzini}, {Sabbadin}, {Simon}, \&
  {Turatto}}]{pcc+07}
{Patat}, F., {Chandra}, P., {Chevalier}, R., {et~al.} 2007, Science, 317, 924

\bibitem[{{Pereira} {et~al.}(2013){Pereira}, {Thomas}, {Aldering}, {Antilogus},
  {Baltay}, {Benitez-Herrera}, {Bongard}, {Buton}, {Canto}, {Cellier-Holzem},
  {Chen}, {Childress}, {Chotard}, {Copin}, {Fakhouri}, {Fink}, {Fouchez},
  {Gangler}, {Guy}, {Hillebrandt}, {Hsiao}, {Kerschhaggl}, {Kowalski},
  {Kromer}, {Nordin}, {Nugent}, {Paech}, {Pain}, {P{\'e}contal}, {Perlmutter},
  {Rabinowitz}, {Rigault}, {Runge}, {Saunders}, {Smadja}, {Tao},
  {Taubenberger}, {Tilquin}, \& {Wu}}]{PTA2013}
{Pereira}, R., {Thomas}, R.~C., {Aldering}, G., {et~al.} 2013, \aap, 554, A27

\bibitem[{{Phillips}(1993)}]{Phillips1993}
{Phillips}, M.~M. 1993, \apjl, 413, L105

\bibitem[{{Phillips} {et~al.}(1999){Phillips}, {Lira}, {Suntzeff}, {Schommer},
  {Hamuy}, \& {Maza}}]{pls+99}
{Phillips}, M.~M., {Lira}, P., {Suntzeff}, N.~B., {et~al.} 1999, \aj, 118, 1766

\bibitem[{{Piro}(2008)}]{p08}
{Piro}, A.~L. 2008, \apj, 679, 616

\bibitem[{{Piro}(2012)}]{p12}
---. 2012, \apj, 759, 83

\bibitem[{{Piro} \& {Morozova}(2015)}]{pm15}
{Piro}, A.~L., \& {Morozova}, V.~S. 2015, ArXiv e-prints, arXiv:1512.03442

\bibitem[{{Poznanski} {et~al.}(2012){Poznanski}, {Prochaska}, \&
  {Bloom}}]{PPB2012}
{Poznanski}, D., {Prochaska}, J.~X., \& {Bloom}, J.~S. 2012, \mnras, 426, 1465

\bibitem[{{Rau} {et~al.}(2009){Rau}, {Kulkarni}, {Law}, {Bloom}, {Ciardi},
  {Djorgovski}, {Fox}, {Gal-Yam}, {Grillmair}, {Kasliwal}, {Nugent}, {Ofek},
  {Quimby}, {Reach}, {Shara}, {Bildsten}, {Cenko}, {Drake}, {Filippenko},
  {Helfand}, {Helou}, {Howell}, {Poznanski}, \& {Sullivan}}]{RKL2009}
{Rau}, A., {Kulkarni}, S.~R., {Law}, N.~M., {et~al.} 2009, \pasp, 121, 1334

\bibitem[{{Riess} {et~al.}(2004){Riess}, {Strolger}, {Tonry}, {Tsvetanov},
  {Casertano}, {Ferguson}, {Mobasher}, {Challis}, {Panagia}, {Filippenko},
  {Li}, {Chornock}, {Kirshner}, {Leibundgut}, {Dickinson}, {Koekemoer},
  {Grogin}, \& {Giavalisco}}]{rst+04}
{Riess}, A.~G., {Strolger}, L.-G., {Tonry}, J., {et~al.} 2004, \apjl, 600, L163

\bibitem[{{Riess} {et~al.}(2007){Riess}, {Strolger}, {Casertano}, {Ferguson},
  {Mobasher}, {Gold}, {Challis}, {Filippenko}, {Jha}, {Li}, {Tonry}, {Foley},
  {Kirshner}, {Dickinson}, {MacDonald}, {Eisenstein}, {Livio}, {Younger}, {Xu},
  {Dahl{\'e}n}, \& {Stern}}]{rsc+07}
{Riess}, A.~G., {Strolger}, L.-G., {Casertano}, S., {et~al.} 2007, \apj, 659,
  98

\bibitem[{{Saio} \& {Nomoto}(2004)}]{sn04}
{Saio}, H., \& {Nomoto}, K. 2004, \apj, 615, 444

\bibitem[{{Scalzo} {et~al.}(2012){Scalzo}, {Aldering}, {Antilogus}, {Aragon},
  {Bailey}, {Baltay}, {Bongard}, {Buton}, {Canto}, {Cellier-Holzem},
  {Childress}, {Chotard}, {Copin}, {Fakhouri}, {Gangler}, {Guy}, {Hsiao},
  {Kerschhaggl}, {Kowalski}, {Nugent}, {Paech}, {Pain}, {Pecontal}, {Pereira},
  {Perlmutter}, {Rabinowitz}, {Rigault}, {Runge}, {Smadja}, {Tao}, {Thomas},
  {Weaver}, {Wu}, \& {Nearby Supernova Factory}}]{saa+12}
{Scalzo}, R., {Aldering}, G., {Antilogus}, P., {et~al.} 2012, \apj, 757, 12

\bibitem[{{Scalzo} {et~al.}(2014{\natexlab{a}}){Scalzo}, {Aldering},
  {Antilogus}, {Aragon}, {Bailey}, {Baltay}, {Bongard}, {Buton},
  {Cellier-Holzem}, {Childress}, {Chotard}, {Copin}, {Fakhouri}, {Gangler},
  {Guy}, {Kim}, {Kowalski}, {Kromer}, {Nordin}, {Nugent}, {Paech}, {Pain},
  {Pecontal}, {Pereira}, {Perlmutter}, {Rabinowitz}, {Rigault}, {Runge},
  {Saunders}, {Sim}, {Smadja}, {Tao}, {Taubenberger}, {Thomas}, {Weaver}, \&
  {Nearby Supernova Factory}}]{SAA2014}
---. 2014{\natexlab{a}}, \mnras, 440, 1498

\bibitem[{{Scalzo} {et~al.}(2014{\natexlab{b}}){Scalzo}, {Ruiter}, \&
  {Sim}}]{sra14}
{Scalzo}, R.~A., {Ruiter}, A.~J., \& {Sim}, S.~A. 2014{\natexlab{b}}, \mnras,
  445, 2535

\bibitem[{{Scalzo} {et~al.}(2010){Scalzo}, {Aldering}, {Antilogus}, {Aragon},
  {Bailey}, {Baltay}, {Bongard}, {Buton}, {Childress}, {Chotard}, {Copin},
  {Fakhouri}, {Gal-Yam}, {Gangler}, {Hoyer}, {Kasliwal}, {Loken}, {Nugent},
  {Pain}, {P{\'e}contal}, {Pereira}, {Perlmutter}, {Rabinowitz}, {Rau},
  {Rigaudier}, {Runge}, {Smadja}, {Tao}, {Thomas}, {Weaver}, \& {Wu}}]{saa+10}
{Scalzo}, R.~A., {Aldering}, G., {Antilogus}, P., {et~al.} 2010, \apj, 713,
  1073

\bibitem[{{Scalzo} {et~al.}(2014{\natexlab{c}}){Scalzo}, {Childress}, {Tucker},
  {Yuan}, {Schmidt}, {Brown}, {Contreras}, {Morrell}, {Hsiao}, {Burns},
  {Phillips}, {Campillay}, {Gonzalez}, {Krisciunas}, {Stritzinger}, {Graham},
  {Parrent}, {Valenti}, {Lidman}, {Schaefer}, {Scott}, {Fraser}, {Gal-Yam},
  {Inserra}, {Maguire}, {Smartt}, {Sollerman}, {Sullivan}, {Taddia}, {Yaron},
  {Young}, {Taubenberger}, {Baltay}, {Ellman}, {Feindt}, {Hadjiyska},
  {McKinnon}, {Nugent}, {Rabinowitz}, \& {Walker}}]{sct+14}
{Scalzo}, R.~A., {Childress}, M., {Tucker}, B., {et~al.} 2014{\natexlab{c}},
  \mnras, 445, 30

\bibitem[{{Schlafly} \& {Finkbeiner}(2011)}]{SF2011}
{Schlafly}, E.~F., \& {Finkbeiner}, D.~P. 2011, \apj, 737, 103

\bibitem[{{Shappee} {et~al.}(2013){Shappee}, {Stanek}, {Pogge}, \&
  {Garnavich}}]{ssp+13}
{Shappee}, B.~J., {Stanek}, K.~Z., {Pogge}, R.~W., \& {Garnavich}, P.~M. 2013,
  \apjl, 762, L5

\bibitem[{{Silverman} \& {Filippenko}(2012)}]{sf12}
{Silverman}, J.~M., \& {Filippenko}, A.~V. 2012, \mnras, 425, 1917

\bibitem[{{Silverman} {et~al.}(2011){Silverman}, {Ganeshalingam}, {Li},
  {Filippenko}, {Miller}, \& {Poznanski}}]{sgl+11}
{Silverman}, J.~M., {Ganeshalingam}, M., {Li}, W., {et~al.} 2011, \mnras, 410,
  585

\bibitem[{{Sim} {et~al.}(2013){Sim}, {Seitenzahl}, {Kromer},
  {Ciaraldi-Schoolmann}, {R{\"o}pke}, {Fink}, {Hillebrandt}, {Pakmor},
  {Ruiter}, \& {Taubenberger}}]{ssk+13}
{Sim}, S.~A., {Seitenzahl}, I.~R., {Kromer}, M., {et~al.} 2013, \mnras, 436,
  333

\bibitem[{{Soker} {et~al.}(2014){Soker}, {Garc{\'{\i}}a-Berro}, \&
  {Althaus}}]{sga14}
{Soker}, N., {Garc{\'{\i}}a-Berro}, E., \& {Althaus}, L.~G. 2014, \mnras, 437,
  L66

\bibitem[{{Stanishev} {et~al.}(2015){Stanishev}, {Goobar}, {Amanullah},
  {Bassett}, {Fantaye}, {Garnavich}, {Hlozek}, {Nordin}, {Okouma}, {Ostman},
  {Sako}, {Scalzo}, \& {Smith}}]{sga+15}
{Stanishev}, V., {Goobar}, A., {Amanullah}, R., {et~al.} 2015, ArXiv e-prints,
  arXiv:1505.07707

\bibitem[{{Sternberg} {et~al.}(2014){Sternberg}, {Gal-Yam}, {Simon}, {Patat},
  {Hillebrandt}, {Phillips}, {Foley}, {Thompson}, {Morrell}, {Chomiuk},
  {Soderberg}, {Yong}, {Kraus}, {Herczeg}, {Hsiao}, {Raskutti}, {Cohen},
  {Mazzali}, \& {Nomoto}}]{sgs+14}
{Sternberg}, A., {Gal-Yam}, A., {Simon}, J.~D., {et~al.} 2014, \mnras, 443,
  1849

\bibitem[{{Sullivan} {et~al.}(2011){Sullivan}, {Guy}, {Conley}, {Regnault},
  {Astier}, {Balland}, {Basa}, {Carlberg}, {Fouchez}, {Hardin}, {Hook},
  {Howell}, {Pain}, {Palanque-Delabrouille}, {Perrett}, {Pritchet}, {Rich},
  {Ruhlmann-Kleider}, {Balam}, {Baumont}, {Ellis}, {Fabbro}, {Fakhouri},
  {Fourmanoit}, {Gonz{\'a}lez-Gait{\'a}n}, {Graham}, {Hudson}, {Hsiao},
  {Kronborg}, {Lidman}, {Mourao}, {Neill}, {Perlmutter}, {Ripoche}, {Suzuki},
  \& {Walker}}]{SGC2011}
{Sullivan}, M., {Guy}, J., {Conley}, A., {et~al.} 2011, \apj, 737, 102

\bibitem[{{Swartz} {et~al.}(1995){Swartz}, {Sutherland}, \& {Harkness}}]{ssh95}
{Swartz}, D.~A., {Sutherland}, P.~G., \& {Harkness}, R.~P. 1995, \apj, 446, 766

\bibitem[{{Taubenberger} {et~al.}(2011){Taubenberger}, {Benetti}, {Childress},
  {Pakmor}, {Hachinger}, {Mazzali}, {Stanishev}, {Elias-Rosa}, {Agnoletto},
  {Bufano}, {Ergon}, {Harutyunyan}, {Inserra}, {Kankare}, {Kromer},
  {Navasardyan}, {Nicolas}, {Pastorello}, {Prosperi}, {Salgado}, {Sollerman},
  {Stritzinger}, {Turatto}, {Valenti}, \& {Hillebrandt}}]{tbc+11}
{Taubenberger}, S., {Benetti}, S., {Childress}, M., {et~al.} 2011, \mnras, 412,
  2735

\bibitem[{{Thomas} {et~al.}(2011{\natexlab{a}}){Thomas}, {Nugent}, \&
  {Meza}}]{tnm11}
{Thomas}, R.~C., {Nugent}, P.~E., \& {Meza}, J.~C. 2011{\natexlab{a}}, \pasp,
  123, 237

\bibitem[{{Thomas} {et~al.}(2011{\natexlab{b}}){Thomas}, {Aldering},
  {Antilogus}, {Aragon}, {Bailey}, {Baltay}, {Bongard}, {Buton}, {Canto},
  {Childress}, {Chotard}, {Copin}, {Fakhouri}, {Gangler}, {Hsiao},
  {Kerschhaggl}, {Kowalski}, {Loken}, {Nugent}, {Paech}, {Pain}, {Pecontal},
  {Pereira}, {Perlmutter}, {Rabinowitz}, {Rigault}, {Rubin}, {Runge}, {Scalzo},
  {Smadja}, {Tao}, {Weaver}, {Wu}, {Brown}, {Milne}, \& {Nearby Supernova
  Factory}}]{taa+11}
{Thomas}, R.~C., {Aldering}, G., {Antilogus}, P., {et~al.} 2011{\natexlab{b}},
  \apj, 743, 27

\bibitem[{{Vink{\'o}} {et~al.}(2012){Vink{\'o}}, {S{\'a}rneczky}, {Tak{\'a}ts},
  {Marion}, {Heged{\"u}s}, {B{\'{\i}}r{\'o}}, {Borkovits}, {Szegedi-Elek},
  {Farkas}, {Klagyivik}, {Kiss}, {Kov{\'a}cs}, {P{\'a}l}, {Szak{\'a}ts},
  {Szalai}, {Szalai}, {Szatm{\'a}ry}, {Szing}, {Vida}, \& {Wheeler}}]{vst+12}
{Vink{\'o}}, J., {S{\'a}rneczky}, K., {Tak{\'a}ts}, K., {et~al.} 2012, \aap,
  546, A12

\bibitem[{{Walker} {et~al.}(2007){Walker}, {Mateo}, {Olszewski}, {Gnedin},
  {Wang}, {Sen}, \& {Woodroofe}}]{wmo+07}
{Walker}, M.~G., {Mateo}, M., {Olszewski}, E.~W., {et~al.} 2007, \apjl, 667,
  L53

\bibitem[{{Whelan} \& {Iben}(1973)}]{wi73}
{Whelan}, J., \& {Iben}, Jr., I. 1973, \apj, 186, 1007

\bibitem[{{Wolf} {et~al.}(2016){Wolf}, {D'Andrea}, {Gupta}, {Sako}, {Fischer},
  {Kessler}, {Jha}, {March}, {Scolnic}, {Fischer}, {Campbell}, {Nichol},
  {Olmstead}, {Richmond}, {Schneider}, \& {Smith}}]{wdg+16}
{Wolf}, R.~C., {D'Andrea}, C.~B., {Gupta}, R.~R., {et~al.} 2016, \apj, 821, 115

\bibitem[{{Yamanaka} {et~al.}(2009){Yamanaka}, {Kawabata}, {Kinugasa},
  {Tanaka}, {Imada}, {Maeda}, {Nomoto}, {Arai}, {Chiyonobu}, {Fukazawa},
  {Hashimoto}, {Honda}, {Ikejiri}, {Itoh}, {Kamata}, {Kawai}, {Komatsu},
  {Konishi}, {Kuroda}, {Miyamoto}, {Miyazaki}, {Nagae}, {Nakaya}, {Ohsugi},
  {Omodaka}, {Sakai}, {Sasada}, {Suzuki}, {Taguchi}, {Takahashi}, {Tanaka},
  {Uemura}, {Yamashita}, {Yanagisawa}, \& {Yoshida}}]{ykk+09}
{Yamanaka}, M., {Kawabata}, K.~S., {Kinugasa}, K., {et~al.} 2009, \apjl, 707,
  L118

\bibitem[{{Yaron} \& {Gal-Yam}(2012)}]{YG2012}
{Yaron}, O., \& {Gal-Yam}, A. 2012, \pasp, 124, 668

\bibitem[{{Yoon} \& {Langer}(2004)}]{yl04}
{Yoon}, S.-C., \& {Langer}, N. 2004, \aap, 419, 623

\bibitem[{{Yoon} \& {Langer}(2005)}]{yl05}
---. 2005, \aap, 435, 967

\bibitem[{{Yuan} {et~al.}(2010){Yuan}, {Quimby}, {Wheeler}, {Vink{\'o}},
  {Chatzopoulos}, {Akerlof}, {Kulkarni}, {Miller}, {McKay}, \&
  {Aharonian}}]{yqw+10}
{Yuan}, F., {Quimby}, R.~M., {Wheeler}, J.~C., {et~al.} 2010, \apj, 715, 1338

\bibitem[{{Zhou} {et~al.}(2013){Zhou}, {Wang}, {Zhang}, {Chen}, {Liang},
  {Zhang}, {Zhou}, {Huang}, {sZhao}, {Wang}, {Zhang}, {Balam}, {Graham}, \&
  {Hsiao}}]{CBET3543}
{Zhou}, L., {Wang}, X., {Zhang}, K., {et~al.} 2013, Central Bureau Electronic
  Telegrams, 3543, 1

\end{thebibliography}
\end{document}